\documentclass[aps,prl,twocolumn,amsmath,amssymb,nofootinbib,superscriptaddress,floatfix,reprint,longbibliography]{revtex4-1}
\usepackage[dvips]{graphicx}
\usepackage{latexsym}
\usepackage{amsmath}
\usepackage{amsfonts}
\usepackage{amssymb}
\usepackage{bm}
\usepackage{color}
\usepackage{txfonts}
\usepackage{float}
\usepackage{url}
\usepackage[colorlinks=true, urlcolor=blue, linkcolor=blue, citecolor=blue, pdftex]{hyperref}
\usepackage{ulem}
\begin{document}
	\newcommand{\fig}[2]{\includegraphics[width=#1]{#2}}
	\newcommand{\la}{{\langle}}
	\newcommand{\ra}{{\rangle}}
	\newcommand{\dg}{{\dagger}}
	\newcommand{\upa}{{\uparrow}}
	\newcommand{\dna}{{\downarrow}}
	\newcommand{\ab}{{\alpha\beta}}
	\newcommand{\ias}{{i\alpha\sigma}}
	\newcommand{\ibs}{{i\beta\sigma}}
	\newcommand{\hH}{\hat{H}}
	\newcommand{\hn}{\hat{n}}
	\newcommand{\hc}{{\hat{\chi}}}
	\newcommand{\hU}{{\hat{U}}}
	\newcommand{\hV}{{\hat{V}}}
	\newcommand{\br}{{\bf r}}
	\newcommand{\bk}{{{\bf k}}}
	\newcommand{\bq}{{{\bf q}}}
	\def\gsim{~\rlap{$>$}{\lower 1.0ex\hbox{$\sim$}}}
	\setlength{\unitlength}{1mm}
	\newcommand{{\vhf}}{$\chi^\text{v}_f$}
	\newcommand{{\vhd}}{$\chi^\text{v}_d$}
	\newcommand{{\vpd}}{$\Delta^\text{v}_d$}
	\newcommand{{\ved}}{$\epsilon^\text{v}_d$}
	\newcommand{{\vved}}{$\varepsilon^\text{v}_d$}
	\newcommand{{\tr}}{{\rm tr}}
	\newcommand{\pprl}{Phys. Rev. Lett. \ }
	\newcommand{\pprb}{Phys. Rev. {B}}

\title{Mottness in two-dimensional van der Waals Nb$_3$X$_8$ monolayers (X=Cl, Br, and I)}
\author{Yi Zhang}
\affiliation{Department of Physics, Shanghai University, Shanghai 200444, China}

\author{Yuhao Gu}
\affiliation{Beijing National Laboratory for Condensed Matter Physics and Institute of Physics,
	Chinese Academy of Sciences, Beijing 100190, China}

\author{Hongming Weng}
\email{hmweng@iphy.ac.cn}
\affiliation{Beijing National Laboratory for Condensed Matter Physics and Institute of Physics,
	Chinese Academy of Sciences, Beijing 100190, China}

\author{Kun Jiang}
\email{jiangkun@iphy.ac.cn}
\affiliation{Beijing National Laboratory for Condensed Matter Physics and Institute of Physics,
	Chinese Academy of Sciences, Beijing 100190, China}
\affiliation{School of Physical Sciences, University of Chinese Academy of Sciences, Beijing 100190, China}

\author{Jiangping Hu}
\email{jphu@iphy.ac.cn}
\affiliation{Beijing National Laboratory for Condensed Matter Physics and Institute of Physics,
	Chinese Academy of Sciences, Beijing 100190, China}
\affiliation{Kavli Institute of Theoretical Sciences, University of Chinese Academy of Sciences,
	Beijing, 100190, China}

\date{\today}

\begin{abstract}
We investigate strong electron-electron correlation effects on 2-dimensional van der Waals materials Nb$_3$X$_8$ (X=Cl, Br, I). We find that the monolayers Nb$_3$X$_8$ are  ideal systems close to  the strong correlation limit. They can be described by a half-filled single band Hubbard model in which the  ratio between the Hubbard, U, and the bandwidth, W,  U/W $\approx$ 5 $\sim$ 10.
Both Mott and magnetic transitions of the material are calculated by the slave boson mean field theory.
Doping the Mott state,  a $d_{x^2-y^2}+id_{xy}$ superconducting pairing instability is found.  We also construct a tunable bilayer Hubbard system for  two sliding Nb$_3$X$_8$ layers. The bilayer system displays a crossover between the band insulator and Mott insulator.
\end{abstract}
\maketitle

\section{Introduction}
Strongly correlated materials display very rich intriguing physical properties \cite{dagotto05}.  For instance,  the cuprates display antiferromagnetic Mott insulator, pseudogap, non-fermi liquid and many other unexpected phenomena besides their high-temperature superconductivity \cite{lee,anderson}. To uncover the microscopic nature of these phenomena, different theoretical methods and numerical techniques, including   resonating valence bond (RVB) \cite{anderson}, slave-boson mean-field \cite{lee}, Gutzwiller approximation \cite{anderson}, dynamical mean-field theory (DMFT) \cite{dmft}, density matrix renormalization group (DMRG) \cite{dmrg} etc., have been developed. However, despite the tremendous efforts over the last forty years,  strongly correlated models are barely solved.  One of the major difficulties is that  the correlated materials are often complicated by a strong interplay between  spin, orbit, charge and lattice degree of freedoms \cite{book1,book2,dagotto}. For example, all five d-orbitals are intertwined in iron-based superconductors \cite{greene,chen_NSR}. Plain-vanilla systems with minimum parameters are highly desirable in strongly correlated materials.  Twisted bilayer graphenes (TBG) with nearly vanishing bandwidth have been considered as  ideal systems to investigate strong correlation in 2-dimensional (2D) van der Waals (vdW) materials \cite{tbg1,tbg2,lu2019,sharpe2019,mao2019,ker2019,xie2019,choi2019,matthew2019,serlin2020}.  But in reality, the low energy physics of TBG is not simple enough  because of the interplay between layer, spin and valley \cite{tbg_review,balents2020,dai2021}. 

Recently, Na$_3$Cl$_8$ has been shown to be a simple Mott insulator \cite{arpes}. Using angle-resolved photoemission spectroscopy (ARPES), photoluminescence spectroscopy (PL)  and DMFT calculations, a half-filled flat band with U around 0.8 eV$\sim$1.2 eV was observed in Na$_3$Cl$_8$ \cite{arpes}.
Additionally, a Josephson junction using Na$_3$Br$_8$ thin flake as a barrier also leads to a long-term chasing superconducting diode effect without the magnetic field, which is related to the strongly correlated nature of this material \cite{ali,ando,nagaosa1,nagaosa17,nagaosa18,hu,yuan,fu_2022,yi}. All these findings show the Nb$_3$X$_8$ (X=Cl, Br, I) may be an ideal platform to study correlation physics. 
In this work, we carry out a comprehensive investigation of  the few layers Nb$_3$X$_8$. Using the monolayer  Nb$_3$X$_8$,  Mott insulator and magnetism at half-filling are found and the doping instability towards unconventional superconductivity is discussed. Furthermore, a controllable bilayer Hubbard model is constructed by sliding two-layer Nb$_3$X$_8$, where the crossover between the band insulator and Mott insulator takes place.
\begin{figure}
	\begin{center}
		\fig{3.4in}{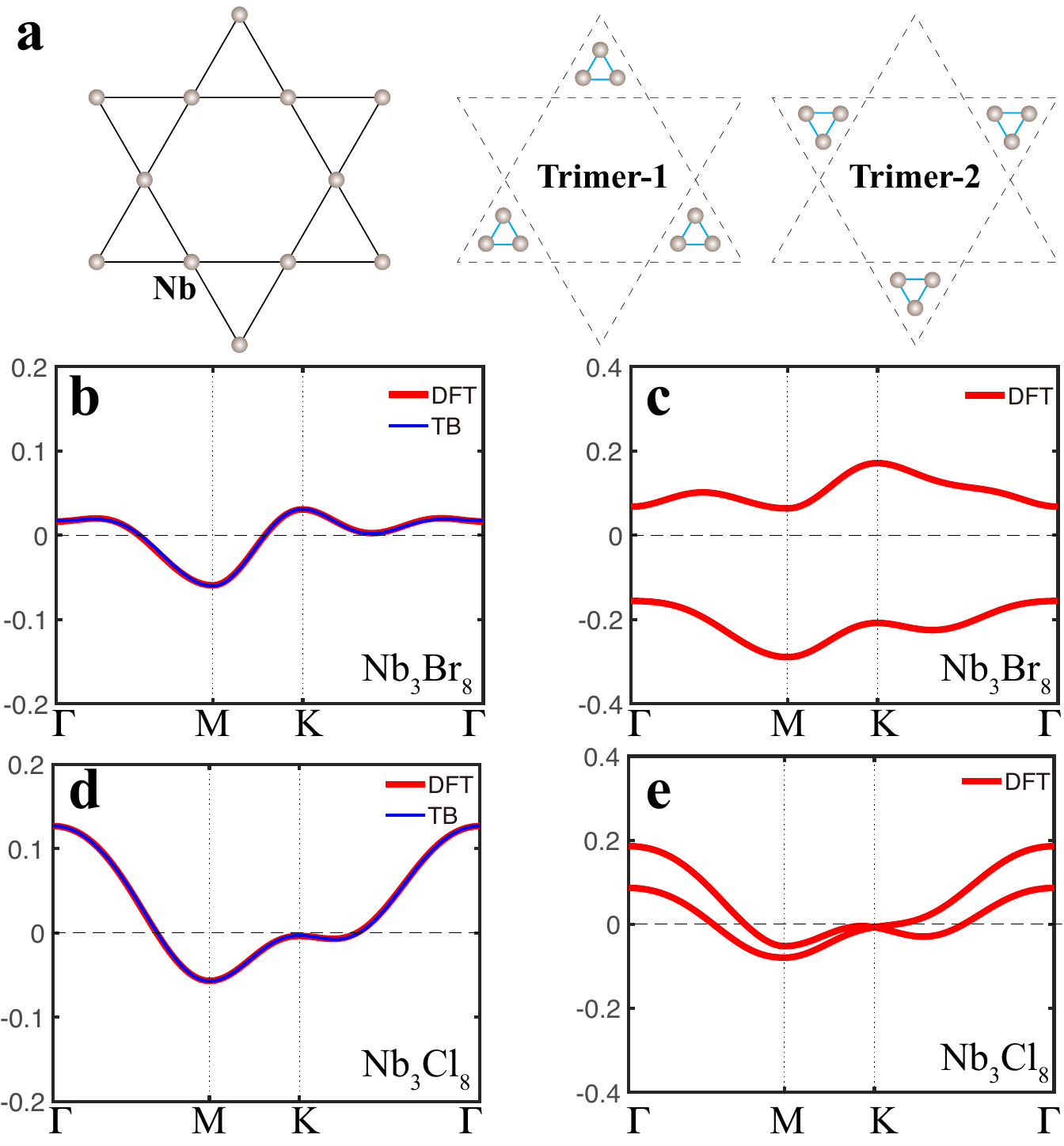}\caption{\textbf{a}, The Nb Kagome lattice dominates the physics of  Nb$_3$X$_8$. The lattice is unstable towards two possible  trimerization structures Trimer-1 and Trimer-2.
			\textbf{b}, The monolayer band structure (in unit of eV) of Nb$_3$Br$_8$ with DFT calculation (red line) and Wannierization fitting TB (blue line). \textbf{c}, The bilayer band structure of Nb$_3$Br$_8$.
			\textbf{d}, The monolayer band structure  of Nb$_3$Cl$_8$. \textbf{e}, The bilayer band structure of Nb$_3$Cl$_8$.
			\label{fig1}}
	\end{center}
	\vskip-0.5cm
\end{figure}

\section{Tight-binding model}
We start from the non-interacting electronic structure of Nb$_3$X$_8$ using density functional theory (DFT) calculation and Wannierization. The crystal structure of  bulk Nb$_3$X$_8$ materials is a standard vdW structure by stacking Nb$_3$X$_8$ along the c direction \cite{schafer,miller,mcqueen17,mcqueen19,nmr,yoon,feng}. 
Owing to the modern development in cleavage technique, reducing vdW materials down to a few layers and monolayer has now become a routine procedure \cite{graphene1,graphene2}. 
Hence, we will focus on the few-layer properties of Nb$_3$X$_8$. Both the bulk and monolayer Nb$_3$X$_8$ electronic structure are dominated by the Nb 4d orbitals, where Nb atoms form a Kagome lattice as shown in Fig.~\ref{fig1}\textbf{a}. However, the Nb Kagome lattice is not a stable structure with instability towards trimerization.
Since the Kagome lattice is formed by two corner-sharing triangles, there are two ways of trimerization, Trimer-1 and Trimer-2, as illustrated in Fig.~\ref{fig1}\textbf{a}. 
Because of this trimerization, the low energy theory of Nb$_3$X$_8$ is changed to a triangle lattice formed by Nb$_3$ clusters. 
For instance, the band structure of monolayer Nb$_3$Br$_8$ around Fermi level is simply one-band, as shown in Fig.~\ref{fig1}\textbf{b}.  
And this band is formed by the 2a$_1$ molecular orbital of Nb 4d$_{z^2}$ trimers \cite{nmr,mcqueen17,arpes}. 
Therefore, the Kagome lattice is reduced to a triangle lattice with a doubled effective lattice constant. Taking into account the in-plane weak hopping of d$_{z^2}$, the bandwidth $W$ is largely reduced down to 0.096 eV.
Similarly, the band structure of monolayer Nb$_3$Cl$_8$ is also showing a one-band character with $W$ around 0.2 eV, as plotted in Fig.~\ref{fig1}\textbf{d}. The band structure of  Nb$_3$I$_8$ is shown in the Supplemental Materials (SM)~\cite{supp} (see also reference ~\onlinecite{kresse1996,Joubert1999,perdew1996}) with a bandwidth around 0.196 eV. Because of the similarity of these systems, we will take Nb$_3$Br$_8$ as an example in the following discussion.

Although the monolayer Nb$_3$Cl$_8$ and Nb$_3$Br$_8$ look similar, their ways of stacking are quite different. In Nb$_3$Br$_8$, each Nb layer is trimerized into Trimer-1 while the Nb$_3$Cl$_8$ layers are trimerized into Trimer-1 and Trimer-2 alternatively. Hence, the Nb$_3$Br$_8$ bilayer is an insulator owing to the strong interlayer hybridization of d$_{z^2}$ orbitals, as plotted in Fig.~\ref{fig1}\textbf{c}. 
On the other hand, the Nb$_3$Cl$_8$ bilayer is weakly coupled owing to the misalignment between two trimers, as shown in Fig.~\ref{fig1}\textbf{e}.

Taking all above facts into account, we can construct the tight-binding (TB) model of Nb$_3$X$_8$ by Wannierization.  The TB model of Nb$_3$X$_8$ monolayer can be written as
\begin{eqnarray}
	H_0= \sum_{ij\sigma} t_{ij} c_{i,\sigma}^{\dagger} c_{j,\sigma}+h.c.-\mu\sum_{i\sigma}c_{i,\sigma}^{\dagger} c_{i,\sigma}
	\label{eq:monolayer}
\end{eqnarray}
Here, we have included the hopping parameters up to the third-nearest-neighbours, with $t_1=5.4$ meV, $t_2=5.7$ meV, $t_3=-6.3$ meV.  As compared in Fig.~\ref{fig1}\textbf{b}, this TB model can faithfully describe the Nb$_3$Br$_8$.
Then, the interaction term is added as the Hubbard model
\begin{eqnarray}
	H_I= U\sum_{i} n_{i\uparrow}n_{i\downarrow}
\end{eqnarray}

To treat Hubbard interaction non-perturbatively and study possible noncollinear magnetism in triangular lattice, we apply the SU(2) spin-rotation invariant Kotliar-Ruckenstein slave boson theory \cite{kr86} and represent the local Hilbert space 
by a spin-1/2 fermion $f_\sigma$ and six bosons $e$, $d$, and $p_\mu$ ($\mu=0,1,2,3$) for empty, doubly-occupied, and singly occupied sites respectively \cite{kr86,li89,wolfle92,jiang14}: $\vert0\rangle=e^\dagger \vert\text{vac}\rangle$,  $|\!\!\uparrow\downarrow\rangle=d^\dagger f_\downarrow^\dagger f_\uparrow^\dagger \vert\text{vac}\rangle$, and $\vert \sigma\rangle= {1\over\sqrt{2}} f_{\sigma^\prime}^\dagger p_\mu^\dagger \tau_{\sigma^\prime\sigma}^\mu \vert \text{vac}\rangle$ where ${\tau}^{1,2,3}$ and ${\tau}^0$ are Pauli and identity matrices \cite{li89,wolfle92,jiang14}. Then, the hopping terms are renormalized to $\psi_{i}^\dagger  g_i^\dagger g_j \psi_j$, where $\psi_i^\dagger=(f_{i\uparrow}^\dagger, f_{i\downarrow}^\dagger)$ and  $g_i$ is the renormalization  factor defined in SM. The Hubbard interaction is written as $U\sum_i {d_i^\dagger  d_i }$ with other local constraints discussed in SM. Then a mean-field solutions are obtained by boson condensation, which are equivalent to the Gutzwiller approximation \cite{wolfle92}.

\begin{figure}
	\begin{center}
		\fig{3.4in}{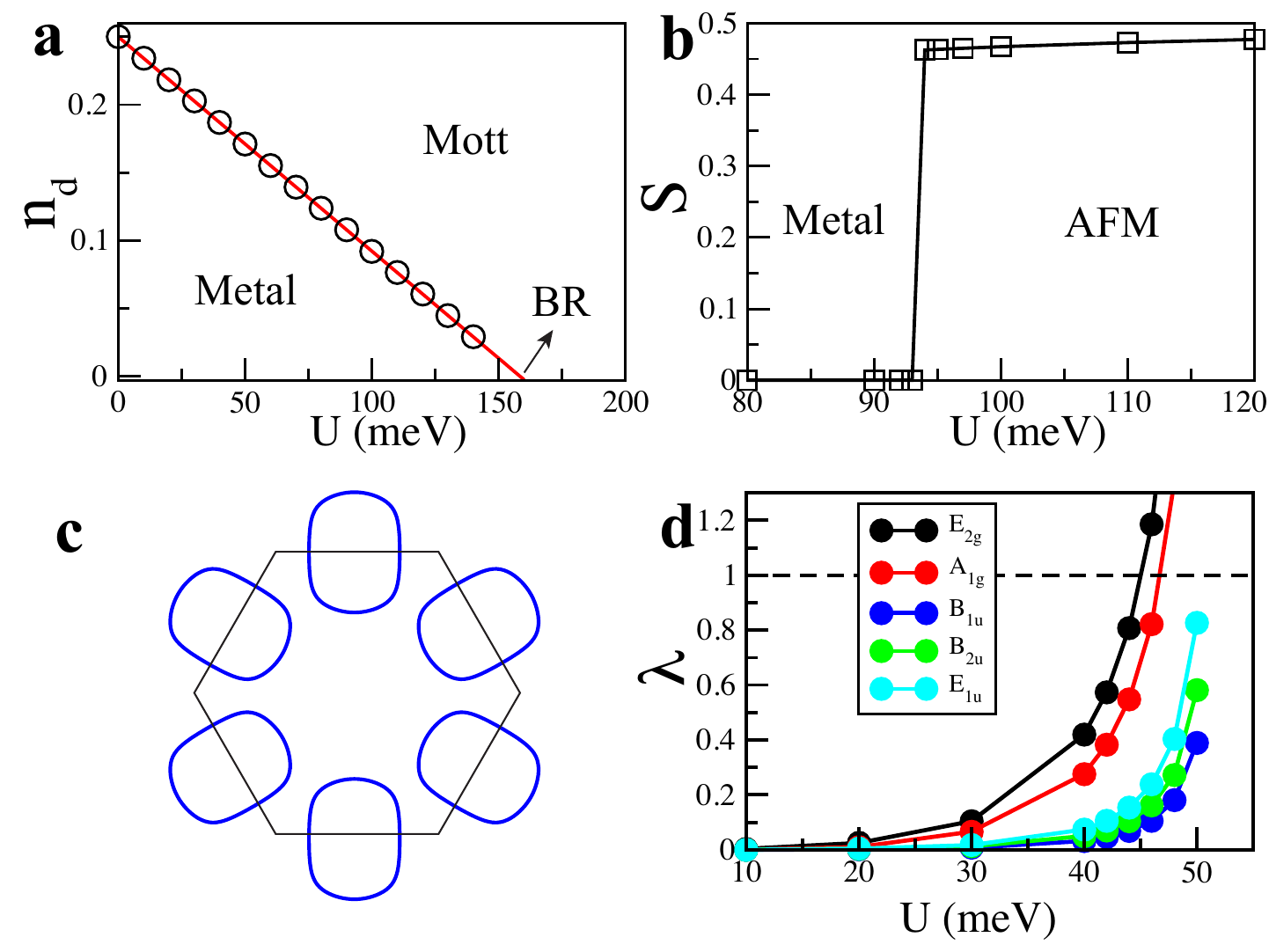}\caption{\textbf{a}, Using slave boson mean field, the paramagnetic solution of monolayer Hubbard model is  determined by the double occupancy $n_d$.
			The BR transition towards Mott occurs at $U_{BR}=152$ meV.
			\textbf{b}, The magnetic expectation value $S$ of monolayer Hubbard model with a phase transition towards $\sqrt{3}\times\sqrt{3}$ AFM at $U=93$ meV.
			\textbf{c}, The Fermi surface of monolayer Nb$_3$Br$_8$ at hole doping $x=0.1$. 
			\textbf{d}, The various SC pairing instabilities determined by RPA approach. The leading instability is the E$_{2g}$ channel (black line) or $d_{x^2-y^2}+id_{xy}$  channel. The subleading one is the A$_{1g}$ s-wave channel (red line).
			\label{fig2}}
	\end{center}
	\vskip-0.5cm
\end{figure}

The paramagnetic solution of the above model can be obtained by setting $p_{x/y/z}$ expectation values to zero. The metal-insulator transition  \cite{mit} is hallmarked by chasing the doublon density $n_d=\langle d_i^\dagger  d_i \rangle$, which is plotted in Fig.~\ref{fig2}\textbf{a}. The $n_d$ linearly decreases as U increases before reaching the Brinkman-Rice (BR) transition \cite{Brinkman} at $U_{BR}=158$ meV, where the renormalization factor $g_i$ vanishes signaling a Mott transition.
Since the Nb Hubbard U is expected to be 0.8$\sim$1.2 eV far beyond $U_{BR}$, the monolayer Nb$_3$X$_8$ lies deep inside the Mott phase. Hence, we need to consider the magnetic solution. 

It is widely known that triangle lattice is a highly frustrated spin system with non-trivial magnetic solutions. The ground state of nearest-neighbor triangle Hubbard model at large U limit and Heisenberg model is $\sqrt{3}\times\sqrt{3}$ antiferromagnetism (AFM) with 120$^\circ$ ordering. 
The large U limit of Nb$_3$Br$_8$ can also be mapped to a highly frustrated $J_1-J_2-J_3$ Heisenberg model. Using slave boson mean-field, a 120$^\circ$ AFM order is found beyond the critical value $U_c=93$ meV, as shown in Fig.\ref{fig2}\textbf{b}. Due to the limitation of mean-field theory, other exotic phases like the quantum spin liquids will be left to future work \cite{zhenyue,becca}.

\section{Random Phase Approximation}
Besides the Mott transition at half-filling, many interesting phenomena can show up by doping a Mott insulator, especially the unconventional superconductivity (SC) . For example,  a very rich phase diagram emerges for a broad doping range in the strongly correlated triangle lattice material Na$_x$CoO$_2$ with unconventional  superconductivity \cite{nacoo1,nacoo2}. This superconductivity is widely believed to emerge from doping a Mott insulator with possible time reversal symmetry breaking \cite{qhwang04,zhouwang07}. The high carrier tunability of 2D materials, through gating and ionic liquids etc., provides a direct way of doping this Mott insulator \cite{graphene2,tbg_review}.
To study the superconductivity instability, we apply a random-phase approximation (RPA) study by hole doping the system \cite{ono,graser}.
Within the RPA approach, the spin susceptibility $\chi^{s}(\bq)$ and charge susceptibility $\chi^{c}(\bq)$ are given as:
\begin{equation}
	\chi^{s/c}(\bq)=\left[1\mp \chi^{0}(\bq)U)\right]^{-1}\chi^{0}(\bq)
\end{equation}
where the $\chi^{0}(\bq)$ is the noninteracting susceptibility.
The effective pairing interaction $V(\bq)$ is constructed as
\begin{equation}
	V(\bq)=\frac{1}{4}\eta U^2 \chi^{s}(\bq)-\frac{1}{8}U^2\chi^{c}(\bq)+\frac{1}{2}U
\end{equation}
with $\eta=\frac{3}{2}$ for spin-singlet state and $\eta=-\frac{1}{2}$ for spin-triplet state.
It governs the superconducting pairing instabilities through the linearized gap $\Delta(\bk)$ equation
\begin{equation}
	\lambda \Delta(\bk)=\frac{1}{N_s}\sum_{k^{\prime}} \frac{f(\varepsilon_{\bk^{\prime}})-f(\varepsilon_{-\bk^{\prime}})}{\varepsilon_{\bk^{\prime}}-(-\varepsilon_{-\bk^{\prime}})} V(\bk-\bk^{\prime})\Delta(\bk^{\prime})
	\label{eq:gap}
\end{equation}
where $N_s$ is the number of sites, $\varepsilon_{\bk}$ is the eigenband of Eq.~\ref{eq:monolayer}, $f(\varepsilon)$ is the Fermi distribution function, and $\lambda$ is the eigenvalue for the gap equation.

\begin{figure}
	\begin{center}
		\fig{3.4in}{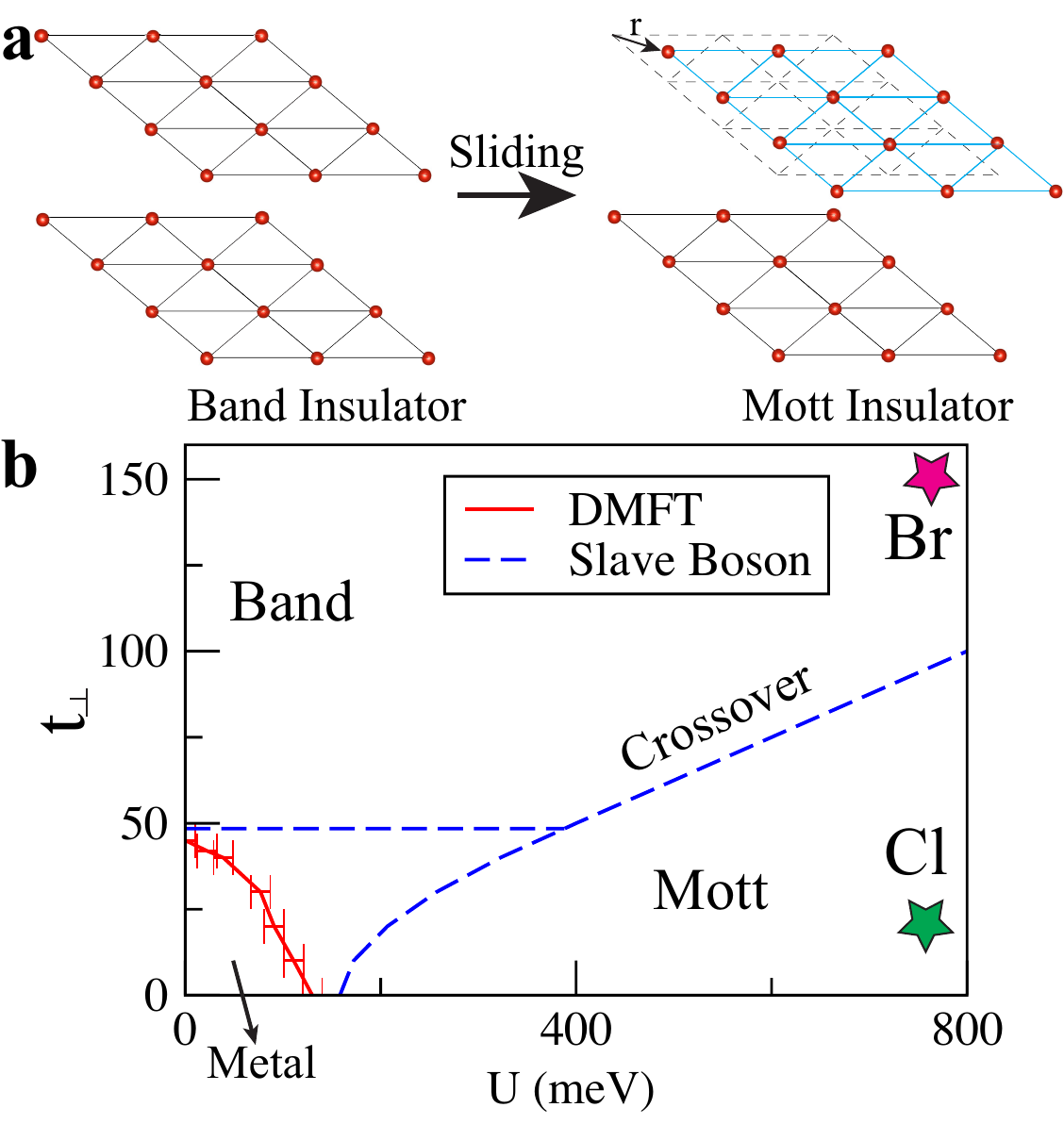}\caption{\textbf{a}, The bilayer Nb$_3$Br$_8$  is used to construct a bilayer Hubbard model. By parallel sliding the top layer with distance $r$, the interlayer coupling $t_{\perp}$ can be accurately adjusted. 
			This sliding approach can be used to study the phase diagram of bilayer Hubbard model.
			\textbf{b}, The phase diagram of bilayer Hubbard model as function of  U  and $t_{\perp}$. The red line is the metal-insulator phase transition line determined by DMFT. There is a crossover between the band insulator limit and Mott insulator limit. We also use the slave boson mean field to find approximate boundary of BR transition (dashed lines), where the metal-insulator  transition line is beyond this approach. The rough positions of bilayer Nb$_3$Br$_8$  and Nb$_3$Cl$_8$  are also labeled by pink star and green star respectively.
			\label{fig3}}
	\end{center}
	\vskip-0.5cm
\end{figure}

We solve the gap equation for the hole dope case with x=0.1, whose Fermi surface is shown in Fig.~\ref{fig3}\textbf{c} and temperature T=1 meV for various values of U within a mesh of 60$\times$60 $\bk$ points.  The dominant pairing symmetry is determined by the gap function $\Delta(\bk)$ whose eigenvalue $\lambda$ become unity firstly. By an approximate $D_{6h}$ point group symmetry, the 
gap functions can be classified as its irreducible representations. In Fig.~\ref{fig2}\textbf{d}, we plot several dominated gap function eigenvalues $\lambda$ as a function of U. The leading SC instability 
is found to be $E_{2g}$ channel, which corresponds to the $d_{x^2-y^2}+id_{xy}$ pairing symmetry. And this $d_{x^2-y^2}+id_{xy}$ pair breaks the time-reversal symmetry with the possible non-trivial topological property \cite{zhouwang07}.  
In the current case considered with hole-doping x=0.1, the system with a next nearest neighbor $d_{x^2-y^2}+id_{xy}$ pairing has Chern number $\pm6$ with more details of the calculation shown in SM~\cite{supp}.
The sub-leading instability is the $A_{1g}$ s-wave and other triplet channels are far away from spin singlet ones.

\section{Bilayer Hubbard model}
Besides the charge tunability, the flexibility of 2D materials and modern electronic techniques make the precise control and engineering of materials and devices possible.  
Successful engineering in small-angle twisted bilayer graphene provides a standard paradigm for this approach, where the bandwidth, interaction screening etc. can be adjusted \cite{tbg1,tbg2,tbg_review}.
Motivated by twisted bilayer graphene, we can also design the bilayer Nb$_3$X$_8$.
By introducing the layer index $l$, the bilayer system can be written as
\begin{equation}
	H=\sum_{l=1,2}  \sum_{ij\sigma} t_{ij} c_{i,l\sigma}^{\dagger} c_{j,l\sigma}  +\sum_{i\sigma}t_{\perp} c_{i,1\sigma}^{\dagger} c_{i,2\sigma}+h.c. + U\sum_{l,i} n_{i,l\uparrow}n_{i,l\downarrow}
	\label{eq:bilayer}
\end{equation}
where the $t_{\perp}$ describes the interlayer coupling. 
Notice that the structure difference in Nb$_3$Cl$_8$ and Nb$_3$Br$_8$ materials leads to a huge difference in $t_{\perp}$ as discussed above.
Therefore, the bilayer system can be tuned by sliding the top layer with distance $r$, as illustrated in Fig.\ref{fig3}\textbf{a}. Then, the $t_{\perp}$  can be approximately written as
\begin{eqnarray}
	t_{\perp}(r)=t_0 \exp(-r/r_0)
\end{eqnarray}
where $r$ is defined as the in-plane distance reference to Nb$_3$Br$_8$ case in Fig.\ref{fig3}\textbf{a}, $t_0=152$ meV is the interlayer coupling strength of Nb$_3$Br$_8$ and
$r_0$ is a decay length by fitting the value at Nb$_3$Cl$_8$, which is around $0.9\sim1.2\AA$.

Hence, the sliding bilayer system is one tunable system by controlling $t_{\perp}$, which can be used to study the bilayer Mott transition.
In the $U=0$ limit, the system undergoes a phase transition from metal to band insulator at large $t_{\perp}$.
On the contrary, if $t_{\perp}$ is zero, there is a Mott transition as discussed in the monolayer system.
One interesting question arises: whether a Mott insulator and a band insulator are fundamentally different?
This question has been widely discussed using both numerical and analytical approaches \cite{kampf03,batista,randeria,scalettar,dagotto}. 
Both crossover and phase transition between them have been identified \cite{ruckenstein,monien,okamoto,kampf,valenti,gall2021}. 
The sliding design here may provide a way to settle down this debate.

Theoretically, we carry out a DMFT study of this model, where the spatial fluctuations of the self-energy are ignored with $\Sigma(k,i\omega_n)\approx \Sigma(i\omega_n)$ \cite{dmft}.
The Hubbard model is further mapped to an Anderson impurity model embedded in an interacting bath with the same self-energy \cite{dmft}. 
Here, we use the hybridization expansion continuous-time quantum Monte Carlo package $i$QIST as the impurity solver \cite{gull,iqist}.

For the bilayer Hubbard model, the self-energy becomes a 2$\times$2 matrix $\hat{\Sigma}(i\omega_n)$. To simplify the impurity calculation, it is more convenient to use the band basis. The band basis $a_{k,\alpha\sigma}$ ($\alpha=\pm$) is easily obtained as
\begin{equation}
			a_{k,\pm\sigma}=\frac{1}{\sqrt{2}}(c_{k,1\sigma}\pm c_{k,2\sigma})
\end{equation}
with eigen-energy $E_{\pm}(k)=\epsilon(k)\pm t_{\perp}$ and $\epsilon(k)$ is monolayer band eigenfunction.
The intereaction terms are transformed into the multi-orbital Hubbard model as
\begin{equation}
	\begin{split}
		H_I&= U_0\sum_{i,\alpha} \hat{n}_{i,\alpha \uparrow}\hat{n}_{i,\alpha \downarrow}+U_v\sum_{i,\alpha\neq \alpha^{\prime}}\hat{n}_{i,\alpha \uparrow}\hat{n}_{i,\alpha^{\prime}\downarrow}
		\\
		& -J\sum_{i,\alpha\ne \alpha^{\prime}} (a_{i,\alpha\uparrow}^{\dagger}a_{i,\alpha^{\prime}\downarrow}^{\dagger}a_{i,\alpha^{\prime}\uparrow}a_{i,\alpha\downarrow} 
	   +a_{i,\alpha\uparrow}^{\dagger}a_{i,\alpha\downarrow}^{\dagger}a_{i,\alpha^{\prime}\uparrow}a_{i,\alpha^{\prime}\downarrow} )
	\end{split}
	\label{eq:HI}
\end{equation}
where $U_0=U_v=J=\frac{U}{2}$.
Then, the self-energy matrix becomes diagonal as
\begin{eqnarray}
	\hat{\Sigma}(i\omega_n)=\left(\begin{array}{cc} 
		\Sigma_{+}(i\omega_n) & 0 \\
		0 & \Sigma_{-}(i\omega_n) 
	\end{array}\right)
\end{eqnarray}

Using DMFT method, the phase diagram of bilayer Hubbard model is obtained, shown in Fig.\ref{fig3}\textbf{b}. 
Besides the $U=0$ transition and  $t_{\perp}$ transition discussed above, a metal-insulator transition line $U_c(t_\perp)$ (red line in Fig.\ref{fig3}(b)) is found by DMFT. 
As shown in Fig.\ref{fig3}\textbf{b}, $U_c(0)=158$ meV as the monolayer value.
However, $U_c(t_\perp)$ keeps decreasing as $t_\perp$ increases before reaching the transition point at  $t_{\perp}\approx$ 48meV and $U=0$.

Besides the metal-insulator transition, no other phase transitions are found from the DMFT calculation. 
The transition between band insulator and Mott insulator is a crossover \cite{okamoto,monien}.
This result shows there is no fundamental difference between band insulator and Mott insulator.
To understand this crossover, we can study the single-site local interaction $\hat{H}_{loc}$ problem with two electrons defined in SM.
The eigenenergy and eigenfunctions for this problem are listed in TABLE \ref{table1}.
\begin{table}[htb]
	\begin{tabular}{|c||c|c|}
		\hline
	    $n$ & Eigenstate $|\Gamma_n \rangle $ & $E_{\Gamma_n}$ \\ \hline \hline
		1 & $|\uparrow \downarrow,0\rangle_a - \zeta |0,\uparrow\downarrow\rangle_a $ & $U/2-\sqrt{16t_{\perp}^2+U^2}/2$ \\ \hline
		2 & $|\uparrow,\downarrow\rangle_a- |\downarrow,\uparrow\rangle_a$& 0\\ \hline
		3 & $|\uparrow,\uparrow\rangle_a$ & 0 \\ \hline
		4 & $|\downarrow,\downarrow\rangle_a$ & 0 \\ \hline
		5 & $|\uparrow,\downarrow\rangle_a+|\downarrow,\uparrow\rangle_a$& U \\ \hline
		6 & $|\uparrow \downarrow,0\rangle_a +\zeta' |0,\uparrow\downarrow\rangle_a $  & $U/2+\sqrt{16t_{\perp}^2+U^2}/2$  \\
		\hline
	\end{tabular}
	\caption{Eigenvectors $|\Gamma_n \rangle $ and their eigenenergy $E_{\Gamma_n}$ for the single-site local interaction problem with two electrons.
	The eigenvectors here are defined in the band basis with subscript a. 
	The first and second components of the state $| \text{1st},\text{2nd}\rangle_a $  stand for the $a_-$ and $a_+$ electrons respectively. We also drop out the normalization factor for convenience. Here, $\zeta=\frac{U}{\sqrt{16t_{\perp}^2+U^2}+4t_{\perp}}$, $\zeta'=\frac{U}{\sqrt{16t_{\perp}^2+U^2}-4t_{\perp}}$.}
		\label{table1}
	\vskip-0.3cm
\end{table}

In TABLE \ref{table1}, we can find the lowest energy state of two electrons is $E_{\Gamma_1}$ with mixing between  $|\uparrow \downarrow,0\rangle_a$ and $|0,\uparrow\downarrow\rangle_a $.
Clearly, in the $U=0$ limit, the ground state of the lattice model is the double occupied the $E_-(k)$ bands formed by the local eigenstate $|\uparrow \downarrow,0\rangle_a$. This state is exactly the lowest energy state  $|\Gamma_1 \rangle $ at $U=0$ limit.
On the other hand, in the large U limit, $|\Gamma_1 \rangle $  will be close to a local bound state $|\uparrow \downarrow,0\rangle_a - |0,\uparrow\downarrow\rangle_a $. 
Switching back to the original c-basis with the two components of $|\ ,\ \rangle_c $ corresponding to the electrons on the two layers respectively, this state becomes $|\uparrow,\downarrow\rangle_c -|\downarrow,\uparrow\rangle_c+\frac{\epsilon}{2t_\perp}(|\uparrow \downarrow,0\rangle_c + |0,\uparrow\downarrow\rangle_c )$ with
$\epsilon=E_{\Gamma_1}$. This approach is similar to the Heitler–London H$_2$ molecule theory, which leads to valence bond theory in chemistry \cite{book1,book2}.
And this bound state forms the Mott insulator in the large U limit. Therefore, a band insulator will smoothly evolve into the Mott insulator without any phase transition \cite{monien,okamoto}. 
More than that, both the band insulator state and Mott insulator bound states are  spin singlet states, which is consistent with experimental results in the bulk Nb$_3$X$_8$ \cite{nmr,mcqueen17}.
Moreover, we calculate the exact retarded Green's function in the atomic limit with more details shown in SM~\cite{supp}. From the pole structure of the Green's function, we see that the interlayer hopping $t_{\perp}$ splits the two Hubbard bands into four which is confirmed by the ARPES results~\cite{arpes}.

In summary, we carry out a comprehensive study of  the correlation physics in monolayer and bilayer Nb$_3$X$_8$. 
The monolayer Nb$_3$X$_8$ is found to be a highly correlated narrow band system with $U/W$ around 5$\sim$10. The Mott transitions are investigated by the slave-boson mean field. Because of the carrier tunability of 2D materials, the superconductivity instability is found to have   $d_{x^2-y^2}+id_{xy}$ pairing.
Additionally, by sliding two Nb$_3$X$_8$ layers, a tunable bilayer Hubbard system is achieved. The bilayer model is further investigated by DMFT, where a crossover between the band insulator and Mott insulator is found.
All these findings show that  the 2-dimensional van der Waals Nb$_3$X$_8$ is  a new platform for correlation physics. We hope our work gives a comprehensive understanding of this material and provides a new route toward correlation effects in 2D.

\begin{acknowledgments}
We thank Tian Qian, Z.-G. Chen for helpful discussions.
This work is supported by the Ministry of Science and Technology  (Grant
No. 2017YFA0303100), National Science Foundation of China (Grant No. NSFC-11888101, No. NSFC-12174428), and the Strategic Priority Research Program of Chinese Academy of Sciences (Grant
No. XDB28000000). 
Y.Z. is supported in part by NSF China Grant No. 12004383, No. 12074276 and No 12274279.
\end{acknowledgments}

\bibliographystyle{apsrev4-1}
\bibliography{reference}

\begin{thebibliography}{72}%
\makeatletter
\providecommand \@ifxundefined [1]{%
 \@ifx{#1\undefined}
}%
\providecommand \@ifnum [1]{%
 \ifnum #1\expandafter \@firstoftwo
 \else \expandafter \@secondoftwo
 \fi
}%
\providecommand \@ifx [1]{%
 \ifx #1\expandafter \@firstoftwo
 \else \expandafter \@secondoftwo
 \fi
}%
\providecommand \natexlab [1]{#1}%
\providecommand \enquote  [1]{``#1''}%
\providecommand \bibnamefont  [1]{#1}%
\providecommand \bibfnamefont [1]{#1}%
\providecommand \citenamefont [1]{#1}%
\providecommand \href@noop [0]{\@secondoftwo}%
\providecommand \href [0]{\begingroup \@sanitize@url \@href}%
\providecommand \@href[1]{\@@startlink{#1}\@@href}%
\providecommand \@@href[1]{\endgroup#1\@@endlink}%
\providecommand \@sanitize@url [0]{\catcode `\\12\catcode `\$12\catcode
  `\&12\catcode `\#12\catcode `\^12\catcode `\_12\catcode `\%12\relax}%
\providecommand \@@startlink[1]{}%
\providecommand \@@endlink[0]{}%
\providecommand \url  [0]{\begingroup\@sanitize@url \@url }%
\providecommand \@url [1]{\endgroup\@href {#1}{\urlprefix }}%
\providecommand \urlprefix  [0]{URL }%
\providecommand \Eprint [0]{\href }%
\providecommand \doibase [0]{http://dx.doi.org/}%
\providecommand \selectlanguage [0]{\@gobble}%
\providecommand \bibinfo  [0]{\@secondoftwo}%
\providecommand \bibfield  [0]{\@secondoftwo}%
\providecommand \translation [1]{[#1]}%
\providecommand \BibitemOpen [0]{}%
\providecommand \bibitemStop [0]{}%
\providecommand \bibitemNoStop [0]{.\EOS\space}%
\providecommand \EOS [0]{\spacefactor3000\relax}%
\providecommand \BibitemShut  [1]{\csname bibitem#1\endcsname}%
\let\auto@bib@innerbib\@empty
\bibitem [{\citenamefont {Dagotto}(2005)}]{dagotto05}%
  \BibitemOpen
  \bibfield  {author} {\bibinfo {author} {\bibfnamefont {E.}~\bibnamefont
  {Dagotto}},\ }\href {\doibase 10.1126/science.1107559} {\bibfield  {journal}
  {\bibinfo  {journal} {Science}\ }\textbf {\bibinfo {volume} {309}},\ \bibinfo
  {pages} {257} (\bibinfo {year} {2005})}\BibitemShut {NoStop}%
\bibitem [{\citenamefont {Lee}\ \emph {et~al.}(2006)\citenamefont {Lee},
  \citenamefont {Nagaosa},\ and\ \citenamefont {Wen}}]{lee}%
  \BibitemOpen
  \bibfield  {author} {\bibinfo {author} {\bibfnamefont {P.~A.}\ \bibnamefont
  {Lee}}, \bibinfo {author} {\bibfnamefont {N.}~\bibnamefont {Nagaosa}}, \ and\
  \bibinfo {author} {\bibfnamefont {X.-G.}\ \bibnamefont {Wen}},\ }\href
  {\doibase 10.1103/RevModPhys.78.17} {\bibfield  {journal} {\bibinfo
  {journal} {Rev. Mod. Phys.}\ }\textbf {\bibinfo {volume} {78}},\ \bibinfo
  {pages} {17} (\bibinfo {year} {2006})}\BibitemShut {NoStop}%
\bibitem [{\citenamefont {Anderson}\ \emph {et~al.}(2004)\citenamefont
  {Anderson}, \citenamefont {Lee}, \citenamefont {Randeria}, \citenamefont
  {Rice}, \citenamefont {Trivedi},\ and\ \citenamefont {Zhang}}]{anderson}%
  \BibitemOpen
  \bibfield  {author} {\bibinfo {author} {\bibfnamefont {P.~W.}\ \bibnamefont
  {Anderson}}, \bibinfo {author} {\bibfnamefont {P.~A.}\ \bibnamefont {Lee}},
  \bibinfo {author} {\bibfnamefont {M.}~\bibnamefont {Randeria}}, \bibinfo
  {author} {\bibfnamefont {T.~M.}\ \bibnamefont {Rice}}, \bibinfo {author}
  {\bibfnamefont {N.}~\bibnamefont {Trivedi}}, \ and\ \bibinfo {author}
  {\bibfnamefont {F.~C.}\ \bibnamefont {Zhang}},\ }\href {\doibase
  10.1088/0953-8984/16/24/r02} {\bibfield  {journal} {\bibinfo  {journal}
  {Journal of Physics: Condensed Matter}\ }\textbf {\bibinfo {volume} {16}},\
  \bibinfo {pages} {R755} (\bibinfo {year} {2004})}\BibitemShut {NoStop}%
\bibitem [{\citenamefont {Georges}\ \emph {et~al.}(1996)\citenamefont
  {Georges}, \citenamefont {Kotliar}, \citenamefont {Krauth},\ and\
  \citenamefont {Rozenberg}}]{dmft}%
  \BibitemOpen
  \bibfield  {author} {\bibinfo {author} {\bibfnamefont {A.}~\bibnamefont
  {Georges}}, \bibinfo {author} {\bibfnamefont {G.}~\bibnamefont {Kotliar}},
  \bibinfo {author} {\bibfnamefont {W.}~\bibnamefont {Krauth}}, \ and\ \bibinfo
  {author} {\bibfnamefont {M.~J.}\ \bibnamefont {Rozenberg}},\ }\href {\doibase
  10.1103/RevModPhys.68.13} {\bibfield  {journal} {\bibinfo  {journal} {Rev.
  Mod. Phys.}\ }\textbf {\bibinfo {volume} {68}},\ \bibinfo {pages} {13}
  (\bibinfo {year} {1996})}\BibitemShut {NoStop}%
\bibitem [{\citenamefont {Schollw\"ock}(2005)}]{dmrg}%
  \BibitemOpen
  \bibfield  {author} {\bibinfo {author} {\bibfnamefont {U.}~\bibnamefont
  {Schollw\"ock}},\ }\href {\doibase 10.1103/RevModPhys.77.259} {\bibfield
  {journal} {\bibinfo  {journal} {Rev. Mod. Phys.}\ }\textbf {\bibinfo {volume}
  {77}},\ \bibinfo {pages} {259} (\bibinfo {year} {2005})}\BibitemShut
  {NoStop}%
\bibitem [{\citenamefont {Fazekas}(1999)}]{book1}%
  \BibitemOpen
  \bibfield  {author} {\bibinfo {author} {\bibfnamefont {P.}~\bibnamefont
  {Fazekas}},\ }\href {\doibase 10.1142/2945} {\emph {\bibinfo {title} {Lecture
  Notes on Electron Correlation and Magnetism}}}\ (\bibinfo  {publisher} {WORLD
  SCIENTIFIC},\ \bibinfo {address} {Singapore},\ \bibinfo {year}
  {1999})\BibitemShut {NoStop}%
\bibitem [{\citenamefont {Auerbach}(1994)}]{book2}%
  \BibitemOpen
  \bibfield  {author} {\bibinfo {author} {\bibfnamefont {A.}~\bibnamefont
  {Auerbach}},\ }\href {\doibase 10.1007/978-1-4612-0869-3} {\emph {\bibinfo
  {title} {Interacting Electrons and Quantum Magnetism}}},\ \bibinfo {edition}
  {1st}\ ed.\ (\bibinfo  {publisher} {Springer New York},\ \bibinfo {address}
  {New York},\ \bibinfo {year} {1994})\BibitemShut {NoStop}%
\bibitem [{\citenamefont {Kancharla}\ and\ \citenamefont
  {Dagotto}(2007)}]{dagotto}%
  \BibitemOpen
  \bibfield  {author} {\bibinfo {author} {\bibfnamefont {S.~S.}\ \bibnamefont
  {Kancharla}}\ and\ \bibinfo {author} {\bibfnamefont {E.}~\bibnamefont
  {Dagotto}},\ }\href {\doibase 10.1103/PhysRevLett.98.016402} {\bibfield
  {journal} {\bibinfo  {journal} {Phys. Rev. Lett.}\ }\textbf {\bibinfo
  {volume} {98}},\ \bibinfo {pages} {016402} (\bibinfo {year}
  {2007})}\BibitemShut {NoStop}%
\bibitem [{\citenamefont {Paglione}\ and\ \citenamefont
  {Greene}(2010)}]{greene}%
  \BibitemOpen
  \bibfield  {author} {\bibinfo {author} {\bibfnamefont {J.}~\bibnamefont
  {Paglione}}\ and\ \bibinfo {author} {\bibfnamefont {R.~L.}\ \bibnamefont
  {Greene}},\ }\href {\doibase 10.1038/nphys1759} {\bibfield  {journal}
  {\bibinfo  {journal} {Nature Physics}\ }\textbf {\bibinfo {volume} {6}},\
  \bibinfo {pages} {645} (\bibinfo {year} {2010})}\BibitemShut {NoStop}%
\bibitem [{\citenamefont {Chen}\ \emph {et~al.}(2014)\citenamefont {Chen},
  \citenamefont {Dai}, \citenamefont {Feng}, \citenamefont {Xiang},\ and\
  \citenamefont {Zhang}}]{chen_NSR}%
  \BibitemOpen
  \bibfield  {author} {\bibinfo {author} {\bibfnamefont {X.}~\bibnamefont
  {Chen}}, \bibinfo {author} {\bibfnamefont {P.}~\bibnamefont {Dai}}, \bibinfo
  {author} {\bibfnamefont {D.}~\bibnamefont {Feng}}, \bibinfo {author}
  {\bibfnamefont {T.}~\bibnamefont {Xiang}}, \ and\ \bibinfo {author}
  {\bibfnamefont {F.-C.}\ \bibnamefont {Zhang}},\ }\href {\doibase
  10.1093/nsr/nwu007} {\bibfield  {journal} {\bibinfo  {journal} {National
  Science Review}\ }\textbf {\bibinfo {volume} {1}},\ \bibinfo {pages} {371}
  (\bibinfo {year} {2014})}\BibitemShut {NoStop}%
\bibitem [{\citenamefont {Cao}\ \emph {et~al.}(2018{\natexlab{a}})\citenamefont
  {Cao}, \citenamefont {Fatemi}, \citenamefont {Demir}, \citenamefont {Fang},
  \citenamefont {Tomarken}, \citenamefont {Luo}, \citenamefont
  {Sanchez-Yamagishi}, \citenamefont {Watanabe}, \citenamefont {Taniguchi},
  \citenamefont {Kaxiras}, \citenamefont {Ashoori},\ and\ \citenamefont
  {Jarillo-Herrero}}]{tbg1}%
  \BibitemOpen
  \bibfield  {author} {\bibinfo {author} {\bibfnamefont {Y.}~\bibnamefont
  {Cao}}, \bibinfo {author} {\bibfnamefont {V.}~\bibnamefont {Fatemi}},
  \bibinfo {author} {\bibfnamefont {A.}~\bibnamefont {Demir}}, \bibinfo
  {author} {\bibfnamefont {S.}~\bibnamefont {Fang}}, \bibinfo {author}
  {\bibfnamefont {S.~L.}\ \bibnamefont {Tomarken}}, \bibinfo {author}
  {\bibfnamefont {J.~Y.}\ \bibnamefont {Luo}}, \bibinfo {author} {\bibfnamefont
  {J.~D.}\ \bibnamefont {Sanchez-Yamagishi}}, \bibinfo {author} {\bibfnamefont
  {K.}~\bibnamefont {Watanabe}}, \bibinfo {author} {\bibfnamefont
  {T.}~\bibnamefont {Taniguchi}}, \bibinfo {author} {\bibfnamefont
  {E.}~\bibnamefont {Kaxiras}}, \bibinfo {author} {\bibfnamefont {R.~C.}\
  \bibnamefont {Ashoori}}, \ and\ \bibinfo {author} {\bibfnamefont
  {P.}~\bibnamefont {Jarillo-Herrero}},\ }\href {\doibase 10.1038/nature26154}
  {\bibfield  {journal} {\bibinfo  {journal} {Nature}\ }\textbf {\bibinfo
  {volume} {556}},\ \bibinfo {pages} {80} (\bibinfo {year}
  {2018}{\natexlab{a}})}\BibitemShut {NoStop}%
\bibitem [{\citenamefont {Cao}\ \emph {et~al.}(2018{\natexlab{b}})\citenamefont
  {Cao}, \citenamefont {Fatemi}, \citenamefont {Fang}, \citenamefont
  {Watanabe}, \citenamefont {Taniguchi}, \citenamefont {Kaxiras},\ and\
  \citenamefont {Jarillo-Herrero}}]{tbg2}%
  \BibitemOpen
  \bibfield  {author} {\bibinfo {author} {\bibfnamefont {Y.}~\bibnamefont
  {Cao}}, \bibinfo {author} {\bibfnamefont {V.}~\bibnamefont {Fatemi}},
  \bibinfo {author} {\bibfnamefont {S.}~\bibnamefont {Fang}}, \bibinfo {author}
  {\bibfnamefont {K.}~\bibnamefont {Watanabe}}, \bibinfo {author}
  {\bibfnamefont {T.}~\bibnamefont {Taniguchi}}, \bibinfo {author}
  {\bibfnamefont {E.}~\bibnamefont {Kaxiras}}, \ and\ \bibinfo {author}
  {\bibfnamefont {P.}~\bibnamefont {Jarillo-Herrero}},\ }\href {\doibase
  10.1038/nature26160} {\bibfield  {journal} {\bibinfo  {journal} {Nature}\
  }\textbf {\bibinfo {volume} {556}},\ \bibinfo {pages} {43} (\bibinfo {year}
  {2018}{\natexlab{b}})}\BibitemShut {NoStop}%
\bibitem [{\citenamefont {Lu}\ \emph {et~al.}(2019)\citenamefont {Lu},
  \citenamefont {Stepanov}, \citenamefont {Yang}, \citenamefont {Xie},
  \citenamefont {Aamir}, \citenamefont {Das}, \citenamefont {Urgell},
  \citenamefont {Watanabe}, \citenamefont {Taniguchi}, \citenamefont {Zhang},
  \citenamefont {Bachtold}, \citenamefont {MacDonald},\ and\ \citenamefont
  {Efetov}}]{lu2019}%
  \BibitemOpen
  \bibfield  {author} {\bibinfo {author} {\bibfnamefont {X.}~\bibnamefont
  {Lu}}, \bibinfo {author} {\bibfnamefont {P.}~\bibnamefont {Stepanov}},
  \bibinfo {author} {\bibfnamefont {W.}~\bibnamefont {Yang}}, \bibinfo {author}
  {\bibfnamefont {M.}~\bibnamefont {Xie}}, \bibinfo {author} {\bibfnamefont
  {M.~A.}\ \bibnamefont {Aamir}}, \bibinfo {author} {\bibfnamefont
  {I.}~\bibnamefont {Das}}, \bibinfo {author} {\bibfnamefont {C.}~\bibnamefont
  {Urgell}}, \bibinfo {author} {\bibfnamefont {K.}~\bibnamefont {Watanabe}},
  \bibinfo {author} {\bibfnamefont {T.}~\bibnamefont {Taniguchi}}, \bibinfo
  {author} {\bibfnamefont {G.}~\bibnamefont {Zhang}}, \bibinfo {author}
  {\bibfnamefont {A.}~\bibnamefont {Bachtold}}, \bibinfo {author}
  {\bibfnamefont {A.~H.}\ \bibnamefont {MacDonald}}, \ and\ \bibinfo {author}
  {\bibfnamefont {D.~K.}\ \bibnamefont {Efetov}},\ }\href {\doibase
  10.1038/s41586-019-1695-0} {\bibfield  {journal} {\bibinfo  {journal}
  {Nature}\ }\textbf {\bibinfo {volume} {574}},\ \bibinfo {pages} {653}
  (\bibinfo {year} {2019})}\BibitemShut {NoStop}%
\bibitem [{\citenamefont {Sharpe}\ \emph {et~al.}(2019)\citenamefont {Sharpe},
  \citenamefont {Fox}, \citenamefont {Barnard}, \citenamefont {Finney},
  \citenamefont {Watanabe}, \citenamefont {Taniguchi}, \citenamefont
  {Kastner},\ and\ \citenamefont {Goldhaber-Gordon}}]{sharpe2019}%
  \BibitemOpen
  \bibfield  {author} {\bibinfo {author} {\bibfnamefont {A.~L.}\ \bibnamefont
  {Sharpe}}, \bibinfo {author} {\bibfnamefont {E.~J.}\ \bibnamefont {Fox}},
  \bibinfo {author} {\bibfnamefont {A.~W.}\ \bibnamefont {Barnard}}, \bibinfo
  {author} {\bibfnamefont {J.}~\bibnamefont {Finney}}, \bibinfo {author}
  {\bibfnamefont {K.}~\bibnamefont {Watanabe}}, \bibinfo {author}
  {\bibfnamefont {T.}~\bibnamefont {Taniguchi}}, \bibinfo {author}
  {\bibfnamefont {M.~A.}\ \bibnamefont {Kastner}}, \ and\ \bibinfo {author}
  {\bibfnamefont {D.}~\bibnamefont {Goldhaber-Gordon}},\ }\href {\doibase
  10.1126/science.aaw3780} {\bibfield  {journal} {\bibinfo  {journal}
  {Science}\ }\textbf {\bibinfo {volume} {365}},\ \bibinfo {pages} {605}
  (\bibinfo {year} {2019})}\BibitemShut {NoStop}%
\bibitem [{\citenamefont {Jiang}\ \emph {et~al.}(2019)\citenamefont {Jiang},
  \citenamefont {Lai}, \citenamefont {Watanabe}, \citenamefont {Taniguchi},
  \citenamefont {Haule}, \citenamefont {Mao},\ and\ \citenamefont
  {Andrei}}]{mao2019}%
  \BibitemOpen
  \bibfield  {author} {\bibinfo {author} {\bibfnamefont {Y.}~\bibnamefont
  {Jiang}}, \bibinfo {author} {\bibfnamefont {X.}~\bibnamefont {Lai}}, \bibinfo
  {author} {\bibfnamefont {K.}~\bibnamefont {Watanabe}}, \bibinfo {author}
  {\bibfnamefont {T.}~\bibnamefont {Taniguchi}}, \bibinfo {author}
  {\bibfnamefont {K.}~\bibnamefont {Haule}}, \bibinfo {author} {\bibfnamefont
  {J.}~\bibnamefont {Mao}}, \ and\ \bibinfo {author} {\bibfnamefont {E.~Y.}\
  \bibnamefont {Andrei}},\ }\href {\doibase 10.1038/s41586-019-1460-4}
  {\bibfield  {journal} {\bibinfo  {journal} {Nature}\ }\textbf {\bibinfo
  {volume} {573}},\ \bibinfo {pages} {91} (\bibinfo {year} {2019})}\BibitemShut
  {NoStop}%
\bibitem [{\citenamefont {Kerelsky}\ \emph {et~al.}(2019)\citenamefont
  {Kerelsky}, \citenamefont {McGilly}, \citenamefont {Kennes}, \citenamefont
  {Xian}, \citenamefont {Yankowitz}, \citenamefont {Chen}, \citenamefont
  {Watanabe}, \citenamefont {Taniguchi}, \citenamefont {Hone}, \citenamefont
  {Dean}, \citenamefont {Rubio},\ and\ \citenamefont {Pasupathy}}]{ker2019}%
  \BibitemOpen
  \bibfield  {author} {\bibinfo {author} {\bibfnamefont {A.}~\bibnamefont
  {Kerelsky}}, \bibinfo {author} {\bibfnamefont {L.~J.}\ \bibnamefont
  {McGilly}}, \bibinfo {author} {\bibfnamefont {D.~M.}\ \bibnamefont {Kennes}},
  \bibinfo {author} {\bibfnamefont {L.}~\bibnamefont {Xian}}, \bibinfo {author}
  {\bibfnamefont {M.}~\bibnamefont {Yankowitz}}, \bibinfo {author}
  {\bibfnamefont {S.}~\bibnamefont {Chen}}, \bibinfo {author} {\bibfnamefont
  {K.}~\bibnamefont {Watanabe}}, \bibinfo {author} {\bibfnamefont
  {T.}~\bibnamefont {Taniguchi}}, \bibinfo {author} {\bibfnamefont
  {J.}~\bibnamefont {Hone}}, \bibinfo {author} {\bibfnamefont {C.}~\bibnamefont
  {Dean}}, \bibinfo {author} {\bibfnamefont {A.}~\bibnamefont {Rubio}}, \ and\
  \bibinfo {author} {\bibfnamefont {A.~N.}\ \bibnamefont {Pasupathy}},\ }\href
  {\doibase 10.1038/s41586-019-1431-9} {\bibfield  {journal} {\bibinfo
  {journal} {Nature}\ }\textbf {\bibinfo {volume} {572}},\ \bibinfo {pages}
  {95} (\bibinfo {year} {2019})}\BibitemShut {NoStop}%
\bibitem [{\citenamefont {Xie}\ \emph {et~al.}(2019)\citenamefont {Xie},
  \citenamefont {Lian}, \citenamefont {J{\"a}ck}, \citenamefont {Liu},
  \citenamefont {Chiu}, \citenamefont {Watanabe}, \citenamefont {Taniguchi},
  \citenamefont {Bernevig},\ and\ \citenamefont {Yazdani}}]{xie2019}%
  \BibitemOpen
  \bibfield  {author} {\bibinfo {author} {\bibfnamefont {Y.}~\bibnamefont
  {Xie}}, \bibinfo {author} {\bibfnamefont {B.}~\bibnamefont {Lian}}, \bibinfo
  {author} {\bibfnamefont {B.}~\bibnamefont {J{\"a}ck}}, \bibinfo {author}
  {\bibfnamefont {X.}~\bibnamefont {Liu}}, \bibinfo {author} {\bibfnamefont
  {C.-L.}\ \bibnamefont {Chiu}}, \bibinfo {author} {\bibfnamefont
  {K.}~\bibnamefont {Watanabe}}, \bibinfo {author} {\bibfnamefont
  {T.}~\bibnamefont {Taniguchi}}, \bibinfo {author} {\bibfnamefont {B.~A.}\
  \bibnamefont {Bernevig}}, \ and\ \bibinfo {author} {\bibfnamefont
  {A.}~\bibnamefont {Yazdani}},\ }\href {\doibase 10.1038/s41586-019-1422-x}
  {\bibfield  {journal} {\bibinfo  {journal} {Nature}\ }\textbf {\bibinfo
  {volume} {572}},\ \bibinfo {pages} {101} (\bibinfo {year}
  {2019})}\BibitemShut {NoStop}%
\bibitem [{\citenamefont {Choi}\ \emph {et~al.}(2019)\citenamefont {Choi},
  \citenamefont {Kemmer}, \citenamefont {Peng}, \citenamefont {Thomson},
  \citenamefont {Arora}, \citenamefont {Polski}, \citenamefont {Zhang},
  \citenamefont {Ren}, \citenamefont {Alicea}, \citenamefont {Refael},
  \citenamefont {von Oppen}, \citenamefont {Watanabe}, \citenamefont
  {Taniguchi},\ and\ \citenamefont {Nadj-Perge}}]{choi2019}%
  \BibitemOpen
  \bibfield  {author} {\bibinfo {author} {\bibfnamefont {Y.}~\bibnamefont
  {Choi}}, \bibinfo {author} {\bibfnamefont {J.}~\bibnamefont {Kemmer}},
  \bibinfo {author} {\bibfnamefont {Y.}~\bibnamefont {Peng}}, \bibinfo {author}
  {\bibfnamefont {A.}~\bibnamefont {Thomson}}, \bibinfo {author} {\bibfnamefont
  {H.}~\bibnamefont {Arora}}, \bibinfo {author} {\bibfnamefont
  {R.}~\bibnamefont {Polski}}, \bibinfo {author} {\bibfnamefont
  {Y.}~\bibnamefont {Zhang}}, \bibinfo {author} {\bibfnamefont
  {H.}~\bibnamefont {Ren}}, \bibinfo {author} {\bibfnamefont {J.}~\bibnamefont
  {Alicea}}, \bibinfo {author} {\bibfnamefont {G.}~\bibnamefont {Refael}},
  \bibinfo {author} {\bibfnamefont {F.}~\bibnamefont {von Oppen}}, \bibinfo
  {author} {\bibfnamefont {K.}~\bibnamefont {Watanabe}}, \bibinfo {author}
  {\bibfnamefont {T.}~\bibnamefont {Taniguchi}}, \ and\ \bibinfo {author}
  {\bibfnamefont {S.}~\bibnamefont {Nadj-Perge}},\ }\href {\doibase
  10.1038/s41567-019-0606-5} {\bibfield  {journal} {\bibinfo  {journal} {Nature
  Physics}\ }\textbf {\bibinfo {volume} {15}},\ \bibinfo {pages} {1174}
  (\bibinfo {year} {2019})}\BibitemShut {NoStop}%
\bibitem [{\citenamefont {Yankowitz}\ \emph {et~al.}(2019)\citenamefont
  {Yankowitz}, \citenamefont {Chen}, \citenamefont {Polshyn}, \citenamefont
  {Zhang}, \citenamefont {Watanabe}, \citenamefont {Taniguchi}, \citenamefont
  {Graf}, \citenamefont {Young},\ and\ \citenamefont {Dean}}]{matthew2019}%
  \BibitemOpen
  \bibfield  {author} {\bibinfo {author} {\bibfnamefont {M.}~\bibnamefont
  {Yankowitz}}, \bibinfo {author} {\bibfnamefont {S.}~\bibnamefont {Chen}},
  \bibinfo {author} {\bibfnamefont {H.}~\bibnamefont {Polshyn}}, \bibinfo
  {author} {\bibfnamefont {Y.}~\bibnamefont {Zhang}}, \bibinfo {author}
  {\bibfnamefont {K.}~\bibnamefont {Watanabe}}, \bibinfo {author}
  {\bibfnamefont {T.}~\bibnamefont {Taniguchi}}, \bibinfo {author}
  {\bibfnamefont {D.}~\bibnamefont {Graf}}, \bibinfo {author} {\bibfnamefont
  {A.~F.}\ \bibnamefont {Young}}, \ and\ \bibinfo {author} {\bibfnamefont
  {C.~R.}\ \bibnamefont {Dean}},\ }\href {\doibase 10.1126/science.aav1910}
  {\bibfield  {journal} {\bibinfo  {journal} {Science}\ }\textbf {\bibinfo
  {volume} {363}},\ \bibinfo {pages} {1059} (\bibinfo {year}
  {2019})}\BibitemShut {NoStop}%
\bibitem [{\citenamefont {Serlin}\ \emph {et~al.}(2020)\citenamefont {Serlin},
  \citenamefont {Tschirhart}, \citenamefont {Polshyn}, \citenamefont {Zhang},
  \citenamefont {Zhu}, \citenamefont {Watanabe}, \citenamefont {Taniguchi},
  \citenamefont {Balents},\ and\ \citenamefont {Young}}]{serlin2020}%
  \BibitemOpen
  \bibfield  {author} {\bibinfo {author} {\bibfnamefont {M.}~\bibnamefont
  {Serlin}}, \bibinfo {author} {\bibfnamefont {C.~L.}\ \bibnamefont
  {Tschirhart}}, \bibinfo {author} {\bibfnamefont {H.}~\bibnamefont {Polshyn}},
  \bibinfo {author} {\bibfnamefont {Y.}~\bibnamefont {Zhang}}, \bibinfo
  {author} {\bibfnamefont {J.}~\bibnamefont {Zhu}}, \bibinfo {author}
  {\bibfnamefont {K.}~\bibnamefont {Watanabe}}, \bibinfo {author}
  {\bibfnamefont {T.}~\bibnamefont {Taniguchi}}, \bibinfo {author}
  {\bibfnamefont {L.}~\bibnamefont {Balents}}, \ and\ \bibinfo {author}
  {\bibfnamefont {A.~F.}\ \bibnamefont {Young}},\ }\href {\doibase
  10.1126/science.aay5533} {\bibfield  {journal} {\bibinfo  {journal}
  {Science}\ }\textbf {\bibinfo {volume} {367}},\ \bibinfo {pages} {900}
  (\bibinfo {year} {2020})}\BibitemShut {NoStop}%
\bibitem [{\citenamefont {Andrei}\ and\ \citenamefont
  {MacDonald}(2020)}]{tbg_review}%
  \BibitemOpen
  \bibfield  {author} {\bibinfo {author} {\bibfnamefont {E.~Y.}\ \bibnamefont
  {Andrei}}\ and\ \bibinfo {author} {\bibfnamefont {A.~H.}\ \bibnamefont
  {MacDonald}},\ }\href {\doibase 10.1038/s41563-020-00840-0} {\bibfield
  {journal} {\bibinfo  {journal} {Nature Materials}\ }\textbf {\bibinfo
  {volume} {19}},\ \bibinfo {pages} {1265} (\bibinfo {year}
  {2020})}\BibitemShut {NoStop}%
\bibitem [{\citenamefont {Balents}\ \emph {et~al.}(2020)\citenamefont
  {Balents}, \citenamefont {Dean}, \citenamefont {Efetov},\ and\ \citenamefont
  {Young}}]{balents2020}%
  \BibitemOpen
  \bibfield  {author} {\bibinfo {author} {\bibfnamefont {L.}~\bibnamefont
  {Balents}}, \bibinfo {author} {\bibfnamefont {C.~R.}\ \bibnamefont {Dean}},
  \bibinfo {author} {\bibfnamefont {D.~K.}\ \bibnamefont {Efetov}}, \ and\
  \bibinfo {author} {\bibfnamefont {A.~F.}\ \bibnamefont {Young}},\ }\href
  {\doibase 10.1038/s41567-020-0906-9} {\bibfield  {journal} {\bibinfo
  {journal} {Nature Physics}\ }\textbf {\bibinfo {volume} {16}},\ \bibinfo
  {pages} {725} (\bibinfo {year} {2020})}\BibitemShut {NoStop}%
\bibitem [{\citenamefont {Liu}\ and\ \citenamefont {Dai}(2021)}]{dai2021}%
  \BibitemOpen
  \bibfield  {author} {\bibinfo {author} {\bibfnamefont {J.}~\bibnamefont
  {Liu}}\ and\ \bibinfo {author} {\bibfnamefont {X.}~\bibnamefont {Dai}},\
  }\href {\doibase 10.1038/s42254-021-00297-3} {\bibfield  {journal} {\bibinfo
  {journal} {Nature Reviews Physics}\ }\textbf {\bibinfo {volume} {3}},\
  \bibinfo {pages} {367} (\bibinfo {year} {2021})}\BibitemShut {NoStop}%
\bibitem [{\citenamefont {{Gao}}\ \emph {et~al.}(2022)\citenamefont {{Gao}},
  \citenamefont {{Zhang}}, \citenamefont {{Wang}}, \citenamefont {{Tao}},
  \citenamefont {{Liu}}, \citenamefont {{Wang}}, \citenamefont {{Yuan}},
  \citenamefont {{Qu}}, \citenamefont {{Pan}}, \citenamefont {{Peng}},
  \citenamefont {{Hu}}, \citenamefont {{Li}}, \citenamefont {{Huang}},
  \citenamefont {{Zhou}}, \citenamefont {{Meng}}, \citenamefont {{Yang}},
  \citenamefont {{Wang}}, \citenamefont {{Yao}}, \citenamefont {{Chen}},
  \citenamefont {{Shi}}, \citenamefont {{Ding}}, \citenamefont {{Jiang}},
  \citenamefont {{Li}}, \citenamefont {{Shi}}, \citenamefont {{Weng}},\ and\
  \citenamefont {{Qian}}}]{arpes}%
  \BibitemOpen
  \bibfield  {author} {\bibinfo {author} {\bibfnamefont {S.}~\bibnamefont
  {{Gao}}}, \bibinfo {author} {\bibfnamefont {S.}~\bibnamefont {{Zhang}}},
  \bibinfo {author} {\bibfnamefont {C.}~\bibnamefont {{Wang}}}, \bibinfo
  {author} {\bibfnamefont {W.}~\bibnamefont {{Tao}}}, \bibinfo {author}
  {\bibfnamefont {J.}~\bibnamefont {{Liu}}}, \bibinfo {author} {\bibfnamefont
  {T.}~\bibnamefont {{Wang}}}, \bibinfo {author} {\bibfnamefont
  {S.}~\bibnamefont {{Yuan}}}, \bibinfo {author} {\bibfnamefont
  {G.}~\bibnamefont {{Qu}}}, \bibinfo {author} {\bibfnamefont {M.}~\bibnamefont
  {{Pan}}}, \bibinfo {author} {\bibfnamefont {S.}~\bibnamefont {{Peng}}},
  \bibinfo {author} {\bibfnamefont {Y.}~\bibnamefont {{Hu}}}, \bibinfo {author}
  {\bibfnamefont {H.}~\bibnamefont {{Li}}}, \bibinfo {author} {\bibfnamefont
  {Y.}~\bibnamefont {{Huang}}}, \bibinfo {author} {\bibfnamefont
  {H.}~\bibnamefont {{Zhou}}}, \bibinfo {author} {\bibfnamefont
  {S.}~\bibnamefont {{Meng}}}, \bibinfo {author} {\bibfnamefont
  {L.}~\bibnamefont {{Yang}}}, \bibinfo {author} {\bibfnamefont
  {Z.}~\bibnamefont {{Wang}}}, \bibinfo {author} {\bibfnamefont
  {Y.}~\bibnamefont {{Yao}}}, \bibinfo {author} {\bibfnamefont
  {Z.}~\bibnamefont {{Chen}}}, \bibinfo {author} {\bibfnamefont
  {M.}~\bibnamefont {{Shi}}}, \bibinfo {author} {\bibfnamefont
  {H.}~\bibnamefont {{Ding}}}, \bibinfo {author} {\bibfnamefont
  {K.}~\bibnamefont {{Jiang}}}, \bibinfo {author} {\bibfnamefont
  {Y.}~\bibnamefont {{Li}}}, \bibinfo {author} {\bibfnamefont {Y.}~\bibnamefont
  {{Shi}}}, \bibinfo {author} {\bibfnamefont {H.}~\bibnamefont {{Weng}}}, \
  and\ \bibinfo {author} {\bibfnamefont {T.}~\bibnamefont {{Qian}}},\
  }\href@noop {} {\bibfield  {journal} {\bibinfo  {journal} {arXiv e-prints}\
  ,\ \bibinfo {eid} {arXiv:2205.11462}} (\bibinfo {year} {2022})},\ \Eprint
  {http://arxiv.org/abs/2205.11462} {arXiv:2205.11462 [cond-mat.str-el]}
  \BibitemShut {NoStop}%
\bibitem [{\citenamefont {Wu}\ \emph {et~al.}(2022)\citenamefont {Wu},
  \citenamefont {Wang}, \citenamefont {Xu}, \citenamefont {Sivakumar},
  \citenamefont {Pasco}, \citenamefont {Filippozzi}, \citenamefont {Parkin},
  \citenamefont {Zeng}, \citenamefont {McQueen},\ and\ \citenamefont
  {Ali}}]{ali}%
  \BibitemOpen
  \bibfield  {author} {\bibinfo {author} {\bibfnamefont {H.}~\bibnamefont
  {Wu}}, \bibinfo {author} {\bibfnamefont {Y.}~\bibnamefont {Wang}}, \bibinfo
  {author} {\bibfnamefont {Y.}~\bibnamefont {Xu}}, \bibinfo {author}
  {\bibfnamefont {P.~K.}\ \bibnamefont {Sivakumar}}, \bibinfo {author}
  {\bibfnamefont {C.}~\bibnamefont {Pasco}}, \bibinfo {author} {\bibfnamefont
  {U.}~\bibnamefont {Filippozzi}}, \bibinfo {author} {\bibfnamefont {S.~S.~P.}\
  \bibnamefont {Parkin}}, \bibinfo {author} {\bibfnamefont {Y.-J.}\
  \bibnamefont {Zeng}}, \bibinfo {author} {\bibfnamefont {T.}~\bibnamefont
  {McQueen}}, \ and\ \bibinfo {author} {\bibfnamefont {M.~N.}\ \bibnamefont
  {Ali}},\ }\href {\doibase 10.1038/s41586-022-04504-8} {\bibfield  {journal}
  {\bibinfo  {journal} {Nature}\ }\textbf {\bibinfo {volume} {604}},\ \bibinfo
  {pages} {653} (\bibinfo {year} {2022})}\BibitemShut {NoStop}%
\bibitem [{\citenamefont {Ando}\ \emph {et~al.}(2020)\citenamefont {Ando},
  \citenamefont {Miyasaka}, \citenamefont {Li}, \citenamefont {Ishizuka},
  \citenamefont {Arakawa}, \citenamefont {Shiota}, \citenamefont {Moriyama},
  \citenamefont {Yanase},\ and\ \citenamefont {Ono}}]{ando}%
  \BibitemOpen
  \bibfield  {author} {\bibinfo {author} {\bibfnamefont {F.}~\bibnamefont
  {Ando}}, \bibinfo {author} {\bibfnamefont {Y.}~\bibnamefont {Miyasaka}},
  \bibinfo {author} {\bibfnamefont {T.}~\bibnamefont {Li}}, \bibinfo {author}
  {\bibfnamefont {J.}~\bibnamefont {Ishizuka}}, \bibinfo {author}
  {\bibfnamefont {T.}~\bibnamefont {Arakawa}}, \bibinfo {author} {\bibfnamefont
  {Y.}~\bibnamefont {Shiota}}, \bibinfo {author} {\bibfnamefont
  {T.}~\bibnamefont {Moriyama}}, \bibinfo {author} {\bibfnamefont
  {Y.}~\bibnamefont {Yanase}}, \ and\ \bibinfo {author} {\bibfnamefont
  {T.}~\bibnamefont {Ono}},\ }\href {\doibase 10.1038/s41586-020-2590-4}
  {\bibfield  {journal} {\bibinfo  {journal} {Nature}\ }\textbf {\bibinfo
  {volume} {584}},\ \bibinfo {pages} {373} (\bibinfo {year}
  {2020})}\BibitemShut {NoStop}%
\bibitem [{\citenamefont {Misaki}\ and\ \citenamefont
  {Nagaosa}(2021)}]{nagaosa1}%
  \BibitemOpen
  \bibfield  {author} {\bibinfo {author} {\bibfnamefont {K.}~\bibnamefont
  {Misaki}}\ and\ \bibinfo {author} {\bibfnamefont {N.}~\bibnamefont
  {Nagaosa}},\ }\href {\doibase 10.1103/PhysRevB.103.245302} {\bibfield
  {journal} {\bibinfo  {journal} {Phys. Rev. B}\ }\textbf {\bibinfo {volume}
  {103}},\ \bibinfo {pages} {245302} (\bibinfo {year} {2021})}\BibitemShut
  {NoStop}%
\bibitem [{\citenamefont {Wakatsuki}\ \emph {et~al.}(2017)\citenamefont
  {Wakatsuki}, \citenamefont {Saito}, \citenamefont {Hoshino}, \citenamefont
  {Itahashi}, \citenamefont {Ideue}, \citenamefont {Ezawa}, \citenamefont
  {Iwasa},\ and\ \citenamefont {Nagaosa}}]{nagaosa17}%
  \BibitemOpen
  \bibfield  {author} {\bibinfo {author} {\bibfnamefont {R.}~\bibnamefont
  {Wakatsuki}}, \bibinfo {author} {\bibfnamefont {Y.}~\bibnamefont {Saito}},
  \bibinfo {author} {\bibfnamefont {S.}~\bibnamefont {Hoshino}}, \bibinfo
  {author} {\bibfnamefont {Y.~M.}\ \bibnamefont {Itahashi}}, \bibinfo {author}
  {\bibfnamefont {T.}~\bibnamefont {Ideue}}, \bibinfo {author} {\bibfnamefont
  {M.}~\bibnamefont {Ezawa}}, \bibinfo {author} {\bibfnamefont
  {Y.}~\bibnamefont {Iwasa}}, \ and\ \bibinfo {author} {\bibfnamefont
  {N.}~\bibnamefont {Nagaosa}},\ }\href {\doibase 10.1126/sciadv.1602390}
  {\bibfield  {journal} {\bibinfo  {journal} {Science Advances}\ }\textbf
  {\bibinfo {volume} {3}},\ \bibinfo {pages} {e1602390} (\bibinfo {year}
  {2017})}\BibitemShut {NoStop}%
\bibitem [{\citenamefont {Tokura}\ and\ \citenamefont
  {Nagaosa}(2018)}]{nagaosa18}%
  \BibitemOpen
  \bibfield  {author} {\bibinfo {author} {\bibfnamefont {Y.}~\bibnamefont
  {Tokura}}\ and\ \bibinfo {author} {\bibfnamefont {N.}~\bibnamefont
  {Nagaosa}},\ }\href {\doibase 10.1038/s41467-018-05759-4} {\bibfield
  {journal} {\bibinfo  {journal} {Nature Communications}\ }\textbf {\bibinfo
  {volume} {9}},\ \bibinfo {pages} {3740} (\bibinfo {year} {2018})}\BibitemShut
  {NoStop}%
\bibitem [{\citenamefont {Hu}\ \emph {et~al.}(2007)\citenamefont {Hu},
  \citenamefont {Wu},\ and\ \citenamefont {Dai}}]{hu}%
  \BibitemOpen
  \bibfield  {author} {\bibinfo {author} {\bibfnamefont {J.}~\bibnamefont
  {Hu}}, \bibinfo {author} {\bibfnamefont {C.}~\bibnamefont {Wu}}, \ and\
  \bibinfo {author} {\bibfnamefont {X.}~\bibnamefont {Dai}},\ }\href {\doibase
  10.1103/PhysRevLett.99.067004} {\bibfield  {journal} {\bibinfo  {journal}
  {Phys. Rev. Lett.}\ }\textbf {\bibinfo {volume} {99}},\ \bibinfo {pages}
  {067004} (\bibinfo {year} {2007})}\BibitemShut {NoStop}%
\bibitem [{\citenamefont {Yuan}\ and\ \citenamefont {Fu}(2022)}]{yuan}%
  \BibitemOpen
  \bibfield  {author} {\bibinfo {author} {\bibfnamefont {N.~F.~Q.}\
  \bibnamefont {Yuan}}\ and\ \bibinfo {author} {\bibfnamefont {L.}~\bibnamefont
  {Fu}},\ }\href {\doibase 10.1073/pnas.2119548119} {\bibfield  {journal}
  {\bibinfo  {journal} {Proceedings of the National Academy of Sciences}\
  }\textbf {\bibinfo {volume} {119}},\ \bibinfo {pages} {e2119548119} (\bibinfo
  {year} {2022})}\BibitemShut {NoStop}%
\bibitem [{\citenamefont {Davydova}\ \emph {et~al.}(2022)\citenamefont
  {Davydova}, \citenamefont {Prembabu},\ and\ \citenamefont {Fu}}]{fu_2022}%
  \BibitemOpen
  \bibfield  {author} {\bibinfo {author} {\bibfnamefont {M.}~\bibnamefont
  {Davydova}}, \bibinfo {author} {\bibfnamefont {S.}~\bibnamefont {Prembabu}},
  \ and\ \bibinfo {author} {\bibfnamefont {L.}~\bibnamefont {Fu}},\ }\href
  {\doibase 10.1126/sciadv.abo0309} {\bibfield  {journal} {\bibinfo  {journal}
  {Science Advances}\ }\textbf {\bibinfo {volume} {8}},\ \bibinfo {pages}
  {eabo0309} (\bibinfo {year} {2022})}\BibitemShut {NoStop}%
\bibitem [{\citenamefont {Zhang}\ \emph {et~al.}(2022)\citenamefont {Zhang},
  \citenamefont {Gu}, \citenamefont {Li}, \citenamefont {Hu},\ and\
  \citenamefont {Jiang}}]{yi}%
  \BibitemOpen
  \bibfield  {author} {\bibinfo {author} {\bibfnamefont {Y.}~\bibnamefont
  {Zhang}}, \bibinfo {author} {\bibfnamefont {Y.}~\bibnamefont {Gu}}, \bibinfo
  {author} {\bibfnamefont {P.}~\bibnamefont {Li}}, \bibinfo {author}
  {\bibfnamefont {J.}~\bibnamefont {Hu}}, \ and\ \bibinfo {author}
  {\bibfnamefont {K.}~\bibnamefont {Jiang}},\ }\href {\doibase
  10.1103/PhysRevX.12.041013} {\bibfield  {journal} {\bibinfo  {journal} {Phys.
  Rev. X}\ }\textbf {\bibinfo {volume} {12}},\ \bibinfo {pages} {041013}
  (\bibinfo {year} {2022})}\BibitemShut {NoStop}%
\bibitem [{\citenamefont {Schäfer}\ and\ \citenamefont
  {Schnering}(1964)}]{schafer}%
  \BibitemOpen
  \bibfield  {author} {\bibinfo {author} {\bibfnamefont {H.}~\bibnamefont
  {Schäfer}}\ and\ \bibinfo {author} {\bibfnamefont {H.}~\bibnamefont
  {Schnering}},\ }\href {\doibase https://doi.org/10.1002/ange.19640762002}
  {\bibfield  {journal} {\bibinfo  {journal} {Angewandte Chemie}\ }\textbf
  {\bibinfo {volume} {76}},\ \bibinfo {pages} {833} (\bibinfo {year}
  {1964})}\BibitemShut {NoStop}%
\bibitem [{\citenamefont {Miller}(1995)}]{miller}%
  \BibitemOpen
  \bibfield  {author} {\bibinfo {author} {\bibfnamefont {G.~J.}\ \bibnamefont
  {Miller}},\ }\href {\doibase https://doi.org/10.1016/0925-8388(94)01298-V}
  {\bibfield  {journal} {\bibinfo  {journal} {Journal of Alloys and Compounds}\
  }\textbf {\bibinfo {volume} {217}},\ \bibinfo {pages} {5} (\bibinfo {year}
  {1995})}\BibitemShut {NoStop}%
\bibitem [{\citenamefont {Sheckelton}\ \emph {et~al.}(2017)\citenamefont
  {Sheckelton}, \citenamefont {Plumb}, \citenamefont {Trump}, \citenamefont
  {Broholm},\ and\ \citenamefont {McQueen}}]{mcqueen17}%
  \BibitemOpen
  \bibfield  {author} {\bibinfo {author} {\bibfnamefont {J.~P.}\ \bibnamefont
  {Sheckelton}}, \bibinfo {author} {\bibfnamefont {K.~W.}\ \bibnamefont
  {Plumb}}, \bibinfo {author} {\bibfnamefont {B.~A.}\ \bibnamefont {Trump}},
  \bibinfo {author} {\bibfnamefont {C.~L.}\ \bibnamefont {Broholm}}, \ and\
  \bibinfo {author} {\bibfnamefont {T.~M.}\ \bibnamefont {McQueen}},\ }\href
  {\doibase 10.1039/C6QI00470A} {\bibfield  {journal} {\bibinfo  {journal}
  {Inorg. Chem. Front.}\ }\textbf {\bibinfo {volume} {4}},\ \bibinfo {pages}
  {481} (\bibinfo {year} {2017})}\BibitemShut {NoStop}%
\bibitem [{\citenamefont {Pasco}\ \emph {et~al.}(2019)\citenamefont {Pasco},
  \citenamefont {El~Baggari}, \citenamefont {Bianco}, \citenamefont
  {Kourkoutis},\ and\ \citenamefont {McQueen}}]{mcqueen19}%
  \BibitemOpen
  \bibfield  {author} {\bibinfo {author} {\bibfnamefont {C.~M.}\ \bibnamefont
  {Pasco}}, \bibinfo {author} {\bibfnamefont {I.}~\bibnamefont {El~Baggari}},
  \bibinfo {author} {\bibfnamefont {E.}~\bibnamefont {Bianco}}, \bibinfo
  {author} {\bibfnamefont {L.~F.}\ \bibnamefont {Kourkoutis}}, \ and\ \bibinfo
  {author} {\bibfnamefont {T.~M.}\ \bibnamefont {McQueen}},\ }\href {\doibase
  10.1021/acsnano.9b04392} {\bibfield  {journal} {\bibinfo  {journal} {ACS
  Nano}\ }\textbf {\bibinfo {volume} {13}},\ \bibinfo {pages} {9457} (\bibinfo
  {year} {2019})}\BibitemShut {NoStop}%
\bibitem [{\citenamefont {Haraguchi}\ \emph {et~al.}(2017)\citenamefont
  {Haraguchi}, \citenamefont {Michioka}, \citenamefont {Ishikawa},
  \citenamefont {Nakano}, \citenamefont {Yamochi}, \citenamefont {Ueda},\ and\
  \citenamefont {Yoshimura}}]{nmr}%
  \BibitemOpen
  \bibfield  {author} {\bibinfo {author} {\bibfnamefont {Y.}~\bibnamefont
  {Haraguchi}}, \bibinfo {author} {\bibfnamefont {C.}~\bibnamefont {Michioka}},
  \bibinfo {author} {\bibfnamefont {M.}~\bibnamefont {Ishikawa}}, \bibinfo
  {author} {\bibfnamefont {Y.}~\bibnamefont {Nakano}}, \bibinfo {author}
  {\bibfnamefont {H.}~\bibnamefont {Yamochi}}, \bibinfo {author} {\bibfnamefont
  {H.}~\bibnamefont {Ueda}}, \ and\ \bibinfo {author} {\bibfnamefont
  {K.}~\bibnamefont {Yoshimura}},\ }\href {\doibase
  10.1021/acs.inorgchem.6b03028} {\bibfield  {journal} {\bibinfo  {journal}
  {Inorganic Chemistry}\ }\textbf {\bibinfo {volume} {56}},\ \bibinfo {pages}
  {3483} (\bibinfo {year} {2017})}\BibitemShut {NoStop}%
\bibitem [{\citenamefont {Yoon}\ \emph {et~al.}(2020)\citenamefont {Yoon},
  \citenamefont {Lesne}, \citenamefont {Sklarek}, \citenamefont {Sheckelton},
  \citenamefont {Pasco}, \citenamefont {Parkin}, \citenamefont {McQueen},\ and\
  \citenamefont {Ali}}]{yoon}%
  \BibitemOpen
  \bibfield  {author} {\bibinfo {author} {\bibfnamefont {J.}~\bibnamefont
  {Yoon}}, \bibinfo {author} {\bibfnamefont {E.}~\bibnamefont {Lesne}},
  \bibinfo {author} {\bibfnamefont {K.}~\bibnamefont {Sklarek}}, \bibinfo
  {author} {\bibfnamefont {J.}~\bibnamefont {Sheckelton}}, \bibinfo {author}
  {\bibfnamefont {C.}~\bibnamefont {Pasco}}, \bibinfo {author} {\bibfnamefont
  {S.~S.~P.}\ \bibnamefont {Parkin}}, \bibinfo {author} {\bibfnamefont {T.~M.}\
  \bibnamefont {McQueen}}, \ and\ \bibinfo {author} {\bibfnamefont {M.~N.}\
  \bibnamefont {Ali}},\ }\href {\doibase 10.1088/1361-648x/ab832b} {\bibfield
  {journal} {\bibinfo  {journal} {Journal of Physics: Condensed Matter}\
  }\textbf {\bibinfo {volume} {32}},\ \bibinfo {pages} {304004} (\bibinfo
  {year} {2020})}\BibitemShut {NoStop}%
\bibitem [{\citenamefont {Sun}\ \emph {et~al.}(2022)\citenamefont {Sun},
  \citenamefont {Zhou}, \citenamefont {Wang}, \citenamefont {Kumar},
  \citenamefont {Geng}, \citenamefont {Yue}, \citenamefont {Han}, \citenamefont
  {Haraguchi}, \citenamefont {Shimada}, \citenamefont {Cheng}, \citenamefont
  {Chen}, \citenamefont {Shi}, \citenamefont {Wu}, \citenamefont {Meng},\ and\
  \citenamefont {Feng}}]{feng}%
  \BibitemOpen
  \bibfield  {author} {\bibinfo {author} {\bibfnamefont {Z.}~\bibnamefont
  {Sun}}, \bibinfo {author} {\bibfnamefont {H.}~\bibnamefont {Zhou}}, \bibinfo
  {author} {\bibfnamefont {C.}~\bibnamefont {Wang}}, \bibinfo {author}
  {\bibfnamefont {S.}~\bibnamefont {Kumar}}, \bibinfo {author} {\bibfnamefont
  {D.}~\bibnamefont {Geng}}, \bibinfo {author} {\bibfnamefont {S.}~\bibnamefont
  {Yue}}, \bibinfo {author} {\bibfnamefont {X.}~\bibnamefont {Han}}, \bibinfo
  {author} {\bibfnamefont {Y.}~\bibnamefont {Haraguchi}}, \bibinfo {author}
  {\bibfnamefont {K.}~\bibnamefont {Shimada}}, \bibinfo {author} {\bibfnamefont
  {P.}~\bibnamefont {Cheng}}, \bibinfo {author} {\bibfnamefont
  {L.}~\bibnamefont {Chen}}, \bibinfo {author} {\bibfnamefont {Y.}~\bibnamefont
  {Shi}}, \bibinfo {author} {\bibfnamefont {K.}~\bibnamefont {Wu}}, \bibinfo
  {author} {\bibfnamefont {S.}~\bibnamefont {Meng}}, \ and\ \bibinfo {author}
  {\bibfnamefont {B.}~\bibnamefont {Feng}},\ }\href {\doibase
  10.1021/acs.nanolett.2c00778} {\bibfield  {journal} {\bibinfo  {journal}
  {Nano Letters}\ }\textbf {\bibinfo {volume} {22}},\ \bibinfo {pages} {4596}
  (\bibinfo {year} {2022})}\BibitemShut {NoStop}%
\bibitem [{\citenamefont {Novoselov}\ \emph {et~al.}(2004)\citenamefont
  {Novoselov}, \citenamefont {Geim}, \citenamefont {Morozov}, \citenamefont
  {Jiang}, \citenamefont {Zhang}, \citenamefont {Dubonos}, \citenamefont
  {Grigorieva},\ and\ \citenamefont {Firsov}}]{graphene1}%
  \BibitemOpen
  \bibfield  {author} {\bibinfo {author} {\bibfnamefont {K.~S.}\ \bibnamefont
  {Novoselov}}, \bibinfo {author} {\bibfnamefont {A.~K.}\ \bibnamefont {Geim}},
  \bibinfo {author} {\bibfnamefont {S.~V.}\ \bibnamefont {Morozov}}, \bibinfo
  {author} {\bibfnamefont {D.}~\bibnamefont {Jiang}}, \bibinfo {author}
  {\bibfnamefont {Y.}~\bibnamefont {Zhang}}, \bibinfo {author} {\bibfnamefont
  {S.~V.}\ \bibnamefont {Dubonos}}, \bibinfo {author} {\bibfnamefont {I.~V.}\
  \bibnamefont {Grigorieva}}, \ and\ \bibinfo {author} {\bibfnamefont {A.~A.}\
  \bibnamefont {Firsov}},\ }\href {\doibase 10.1126/science.1102896} {\bibfield
   {journal} {\bibinfo  {journal} {Science}\ }\textbf {\bibinfo {volume}
  {306}},\ \bibinfo {pages} {666} (\bibinfo {year} {2004})}\BibitemShut
  {NoStop}%
\bibitem [{\citenamefont {Geim}\ and\ \citenamefont
  {Novoselov}(2007)}]{graphene2}%
  \BibitemOpen
  \bibfield  {author} {\bibinfo {author} {\bibfnamefont {A.~K.}\ \bibnamefont
  {Geim}}\ and\ \bibinfo {author} {\bibfnamefont {K.~S.}\ \bibnamefont
  {Novoselov}},\ }\href {\doibase 10.1038/nmat1849} {\bibfield  {journal}
  {\bibinfo  {journal} {Nature Materials}\ }\textbf {\bibinfo {volume} {6}},\
  \bibinfo {pages} {183} (\bibinfo {year} {2007})}\BibitemShut {NoStop}%
\bibitem [{sup()}]{supp}%
  \BibitemOpen
  \href@noop {} {}\bibinfo {note} {Technical details are provided in the
  Supplemental Materials}\BibitemShut {NoStop}%
\bibitem [{\citenamefont {Kresse}\ and\ \citenamefont
  {Furthmüller}(1996)}]{kresse1996}%
  \BibitemOpen
  \bibfield  {author} {\bibinfo {author} {\bibfnamefont {G.}~\bibnamefont
  {Kresse}}\ and\ \bibinfo {author} {\bibfnamefont {J.}~\bibnamefont
  {Furthmüller}},\ }\href {\doibase
  https://doi.org/10.1016/0927-0256(96)00008-0} {\bibfield  {journal} {\bibinfo
   {journal} {Computational Materials Science}\ }\textbf {\bibinfo {volume}
  {6}},\ \bibinfo {pages} {15} (\bibinfo {year} {1996})}\BibitemShut {NoStop}%
\bibitem [{\citenamefont {Kresse}\ and\ \citenamefont
  {Joubert}(1999)}]{Joubert1999}%
  \BibitemOpen
  \bibfield  {author} {\bibinfo {author} {\bibfnamefont {G.}~\bibnamefont
  {Kresse}}\ and\ \bibinfo {author} {\bibfnamefont {D.}~\bibnamefont
  {Joubert}},\ }\href {\doibase 10.1103/PhysRevB.59.1758} {\bibfield  {journal}
  {\bibinfo  {journal} {Phys. Rev. B}\ }\textbf {\bibinfo {volume} {59}},\
  \bibinfo {pages} {1758} (\bibinfo {year} {1999})}\BibitemShut {NoStop}%
\bibitem [{\citenamefont {Perdew}\ \emph {et~al.}(1996)\citenamefont {Perdew},
  \citenamefont {Burke},\ and\ \citenamefont {Ernzerhof}}]{perdew1996}%
  \BibitemOpen
  \bibfield  {author} {\bibinfo {author} {\bibfnamefont {J.~P.}\ \bibnamefont
  {Perdew}}, \bibinfo {author} {\bibfnamefont {K.}~\bibnamefont {Burke}}, \
  and\ \bibinfo {author} {\bibfnamefont {M.}~\bibnamefont {Ernzerhof}},\ }\href
  {\doibase 10.1103/PhysRevLett.77.3865} {\bibfield  {journal} {\bibinfo
  {journal} {Phys. Rev. Lett.}\ }\textbf {\bibinfo {volume} {77}},\ \bibinfo
  {pages} {3865} (\bibinfo {year} {1996})}\BibitemShut {NoStop}%
\bibitem [{\citenamefont {Kotliar}\ and\ \citenamefont
  {Ruckenstein}(1986)}]{kr86}%
  \BibitemOpen
  \bibfield  {author} {\bibinfo {author} {\bibfnamefont {G.}~\bibnamefont
  {Kotliar}}\ and\ \bibinfo {author} {\bibfnamefont {A.~E.}\ \bibnamefont
  {Ruckenstein}},\ }\href {\doibase 10.1103/PhysRevLett.57.1362} {\bibfield
  {journal} {\bibinfo  {journal} {Phys. Rev. Lett.}\ }\textbf {\bibinfo
  {volume} {57}},\ \bibinfo {pages} {1362} (\bibinfo {year}
  {1986})}\BibitemShut {NoStop}%
\bibitem [{\citenamefont {Li}\ \emph {et~al.}(1989)\citenamefont {Li},
  \citenamefont {W\"olfle},\ and\ \citenamefont {Hirschfeld}}]{li89}%
  \BibitemOpen
  \bibfield  {author} {\bibinfo {author} {\bibfnamefont {T.}~\bibnamefont
  {Li}}, \bibinfo {author} {\bibfnamefont {P.}~\bibnamefont {W\"olfle}}, \ and\
  \bibinfo {author} {\bibfnamefont {P.~J.}\ \bibnamefont {Hirschfeld}},\ }\href
  {\doibase 10.1103/PhysRevB.40.6817} {\bibfield  {journal} {\bibinfo
  {journal} {Phys. Rev. B}\ }\textbf {\bibinfo {volume} {40}},\ \bibinfo
  {pages} {6817} (\bibinfo {year} {1989})}\BibitemShut {NoStop}%
\bibitem [{\citenamefont {Frésard}\ and\ \citenamefont
  {Wölfle}(1992)}]{wolfle92}%
  \BibitemOpen
  \bibfield  {author} {\bibinfo {author} {\bibfnamefont {R.}~\bibnamefont
  {Frésard}}\ and\ \bibinfo {author} {\bibfnamefont {P.}~\bibnamefont
  {Wölfle}},\ }\href {\doibase 10.1142/S0217979292000414} {\bibfield
  {journal} {\bibinfo  {journal} {International Journal of Modern Physics B}\
  }\textbf {\bibinfo {volume} {06}},\ \bibinfo {pages} {685} (\bibinfo {year}
  {1992})}\BibitemShut {NoStop}%
\bibitem [{\citenamefont {Jiang}\ \emph {et~al.}(2014)\citenamefont {Jiang},
  \citenamefont {Zhou},\ and\ \citenamefont {Wang}}]{jiang14}%
  \BibitemOpen
  \bibfield  {author} {\bibinfo {author} {\bibfnamefont {K.}~\bibnamefont
  {Jiang}}, \bibinfo {author} {\bibfnamefont {S.}~\bibnamefont {Zhou}}, \ and\
  \bibinfo {author} {\bibfnamefont {Z.}~\bibnamefont {Wang}},\ }\href {\doibase
  10.1103/PhysRevB.90.165135} {\bibfield  {journal} {\bibinfo  {journal} {Phys.
  Rev. B}\ }\textbf {\bibinfo {volume} {90}},\ \bibinfo {pages} {165135}
  (\bibinfo {year} {2014})}\BibitemShut {NoStop}%
\bibitem [{\citenamefont {Imada}\ \emph {et~al.}(1998)\citenamefont {Imada},
  \citenamefont {Fujimori},\ and\ \citenamefont {Tokura}}]{mit}%
  \BibitemOpen
  \bibfield  {author} {\bibinfo {author} {\bibfnamefont {M.}~\bibnamefont
  {Imada}}, \bibinfo {author} {\bibfnamefont {A.}~\bibnamefont {Fujimori}}, \
  and\ \bibinfo {author} {\bibfnamefont {Y.}~\bibnamefont {Tokura}},\ }\href
  {\doibase 10.1103/RevModPhys.70.1039} {\bibfield  {journal} {\bibinfo
  {journal} {Rev. Mod. Phys.}\ }\textbf {\bibinfo {volume} {70}},\ \bibinfo
  {pages} {1039} (\bibinfo {year} {1998})}\BibitemShut {NoStop}%
\bibitem [{\citenamefont {Brinkman}\ and\ \citenamefont
  {Rice}(1970)}]{Brinkman}%
  \BibitemOpen
  \bibfield  {author} {\bibinfo {author} {\bibfnamefont {W.~F.}\ \bibnamefont
  {Brinkman}}\ and\ \bibinfo {author} {\bibfnamefont {T.~M.}\ \bibnamefont
  {Rice}},\ }\href {\doibase 10.1103/PhysRevB.2.4302} {\bibfield  {journal}
  {\bibinfo  {journal} {Phys. Rev. B}\ }\textbf {\bibinfo {volume} {2}},\
  \bibinfo {pages} {4302} (\bibinfo {year} {1970})}\BibitemShut {NoStop}%
\bibitem [{\citenamefont {Zhu}\ and\ \citenamefont {White}(2015)}]{zhenyue}%
  \BibitemOpen
  \bibfield  {author} {\bibinfo {author} {\bibfnamefont {Z.}~\bibnamefont
  {Zhu}}\ and\ \bibinfo {author} {\bibfnamefont {S.~R.}\ \bibnamefont
  {White}},\ }\href {\doibase 10.1103/PhysRevB.92.041105} {\bibfield  {journal}
  {\bibinfo  {journal} {Phys. Rev. B}\ }\textbf {\bibinfo {volume} {92}},\
  \bibinfo {pages} {041105} (\bibinfo {year} {2015})}\BibitemShut {NoStop}%
\bibitem [{\citenamefont {Iqbal}\ \emph {et~al.}(2016)\citenamefont {Iqbal},
  \citenamefont {Hu}, \citenamefont {Thomale}, \citenamefont {Poilblanc},\ and\
  \citenamefont {Becca}}]{becca}%
  \BibitemOpen
  \bibfield  {author} {\bibinfo {author} {\bibfnamefont {Y.}~\bibnamefont
  {Iqbal}}, \bibinfo {author} {\bibfnamefont {W.-J.}\ \bibnamefont {Hu}},
  \bibinfo {author} {\bibfnamefont {R.}~\bibnamefont {Thomale}}, \bibinfo
  {author} {\bibfnamefont {D.}~\bibnamefont {Poilblanc}}, \ and\ \bibinfo
  {author} {\bibfnamefont {F.}~\bibnamefont {Becca}},\ }\href {\doibase
  10.1103/PhysRevB.93.144411} {\bibfield  {journal} {\bibinfo  {journal} {Phys.
  Rev. B}\ }\textbf {\bibinfo {volume} {93}},\ \bibinfo {pages} {144411}
  (\bibinfo {year} {2016})}\BibitemShut {NoStop}%
\bibitem [{\citenamefont {Takada}\ \emph {et~al.}(2003)\citenamefont {Takada},
  \citenamefont {Sakurai}, \citenamefont {Takayama-Muromachi}, \citenamefont
  {Izumi}, \citenamefont {Dilanian},\ and\ \citenamefont {Sasaki}}]{nacoo1}%
  \BibitemOpen
  \bibfield  {author} {\bibinfo {author} {\bibfnamefont {K.}~\bibnamefont
  {Takada}}, \bibinfo {author} {\bibfnamefont {H.}~\bibnamefont {Sakurai}},
  \bibinfo {author} {\bibfnamefont {E.}~\bibnamefont {Takayama-Muromachi}},
  \bibinfo {author} {\bibfnamefont {F.}~\bibnamefont {Izumi}}, \bibinfo
  {author} {\bibfnamefont {R.~A.}\ \bibnamefont {Dilanian}}, \ and\ \bibinfo
  {author} {\bibfnamefont {T.}~\bibnamefont {Sasaki}},\ }\href {\doibase
  10.1038/nature01450} {\bibfield  {journal} {\bibinfo  {journal} {Nature}\
  }\textbf {\bibinfo {volume} {422}},\ \bibinfo {pages} {53} (\bibinfo {year}
  {2003})}\BibitemShut {NoStop}%
\bibitem [{\citenamefont {Foo}\ \emph {et~al.}(2004)\citenamefont {Foo},
  \citenamefont {Wang}, \citenamefont {Watauchi}, \citenamefont {Zandbergen},
  \citenamefont {He}, \citenamefont {Cava},\ and\ \citenamefont
  {Ong}}]{nacoo2}%
  \BibitemOpen
  \bibfield  {author} {\bibinfo {author} {\bibfnamefont {M.~L.}\ \bibnamefont
  {Foo}}, \bibinfo {author} {\bibfnamefont {Y.}~\bibnamefont {Wang}}, \bibinfo
  {author} {\bibfnamefont {S.}~\bibnamefont {Watauchi}}, \bibinfo {author}
  {\bibfnamefont {H.~W.}\ \bibnamefont {Zandbergen}}, \bibinfo {author}
  {\bibfnamefont {T.}~\bibnamefont {He}}, \bibinfo {author} {\bibfnamefont
  {R.~J.}\ \bibnamefont {Cava}}, \ and\ \bibinfo {author} {\bibfnamefont
  {N.~P.}\ \bibnamefont {Ong}},\ }\href {\doibase
  10.1103/PhysRevLett.92.247001} {\bibfield  {journal} {\bibinfo  {journal}
  {Phys. Rev. Lett.}\ }\textbf {\bibinfo {volume} {92}},\ \bibinfo {pages}
  {247001} (\bibinfo {year} {2004})}\BibitemShut {NoStop}%
\bibitem [{\citenamefont {Wang}\ \emph {et~al.}(2004)\citenamefont {Wang},
  \citenamefont {Lee},\ and\ \citenamefont {Lee}}]{qhwang04}%
  \BibitemOpen
  \bibfield  {author} {\bibinfo {author} {\bibfnamefont {Q.-H.}\ \bibnamefont
  {Wang}}, \bibinfo {author} {\bibfnamefont {D.-H.}\ \bibnamefont {Lee}}, \
  and\ \bibinfo {author} {\bibfnamefont {P.~A.}\ \bibnamefont {Lee}},\ }\href
  {\doibase 10.1103/PhysRevB.69.092504} {\bibfield  {journal} {\bibinfo
  {journal} {Phys. Rev. B}\ }\textbf {\bibinfo {volume} {69}},\ \bibinfo
  {pages} {092504} (\bibinfo {year} {2004})}\BibitemShut {NoStop}%
\bibitem [{\citenamefont {Zhou}\ and\ \citenamefont {Wang}(2008)}]{zhouwang07}%
  \BibitemOpen
  \bibfield  {author} {\bibinfo {author} {\bibfnamefont {S.}~\bibnamefont
  {Zhou}}\ and\ \bibinfo {author} {\bibfnamefont {Z.}~\bibnamefont {Wang}},\
  }\href {\doibase 10.1103/PhysRevLett.100.217002} {\bibfield  {journal}
  {\bibinfo  {journal} {Phys. Rev. Lett.}\ }\textbf {\bibinfo {volume} {100}},\
  \bibinfo {pages} {217002} (\bibinfo {year} {2008})}\BibitemShut {NoStop}%
\bibitem [{\citenamefont {Yanagi}\ \emph {et~al.}(2008)\citenamefont {Yanagi},
  \citenamefont {Yamakawa},\ and\ \citenamefont {Ōno}}]{ono}%
  \BibitemOpen
  \bibfield  {author} {\bibinfo {author} {\bibfnamefont {Y.}~\bibnamefont
  {Yanagi}}, \bibinfo {author} {\bibfnamefont {Y.}~\bibnamefont {Yamakawa}}, \
  and\ \bibinfo {author} {\bibfnamefont {Y.}~\bibnamefont {Ōno}},\ }\href
  {\doibase 10.1143/JPSJ.77.123701} {\bibfield  {journal} {\bibinfo  {journal}
  {Journal of the Physical Society of Japan}\ }\textbf {\bibinfo {volume}
  {77}},\ \bibinfo {pages} {123701} (\bibinfo {year} {2008})}\BibitemShut
  {NoStop}%
\bibitem [{\citenamefont {Graser}\ \emph {et~al.}(2009)\citenamefont {Graser},
  \citenamefont {Maier}, \citenamefont {Hirschfeld},\ and\ \citenamefont
  {Scalapino}}]{graser}%
  \BibitemOpen
  \bibfield  {author} {\bibinfo {author} {\bibfnamefont {S.}~\bibnamefont
  {Graser}}, \bibinfo {author} {\bibfnamefont {T.~A.}\ \bibnamefont {Maier}},
  \bibinfo {author} {\bibfnamefont {P.~J.}\ \bibnamefont {Hirschfeld}}, \ and\
  \bibinfo {author} {\bibfnamefont {D.~J.}\ \bibnamefont {Scalapino}},\ }\href
  {\doibase 10.1088/1367-2630/11/2/025016} {\bibfield  {journal} {\bibinfo
  {journal} {New Journal of Physics}\ }\textbf {\bibinfo {volume} {11}},\
  \bibinfo {pages} {025016} (\bibinfo {year} {2009})}\BibitemShut {NoStop}%
\bibitem [{\citenamefont {Kampf}\ \emph {et~al.}(2003)\citenamefont {Kampf},
  \citenamefont {Sekania}, \citenamefont {Japaridze},\ and\ \citenamefont
  {Brune}}]{kampf03}%
  \BibitemOpen
  \bibfield  {author} {\bibinfo {author} {\bibfnamefont {A.~P.}\ \bibnamefont
  {Kampf}}, \bibinfo {author} {\bibfnamefont {M.}~\bibnamefont {Sekania}},
  \bibinfo {author} {\bibfnamefont {G.~I.}\ \bibnamefont {Japaridze}}, \ and\
  \bibinfo {author} {\bibfnamefont {P.}~\bibnamefont {Brune}},\ }\href
  {\doibase 10.1088/0953-8984/15/34/319} {\bibfield  {journal} {\bibinfo
  {journal} {Journal of Physics: Condensed Matter}\ }\textbf {\bibinfo {volume}
  {15}},\ \bibinfo {pages} {5895} (\bibinfo {year} {2003})}\BibitemShut
  {NoStop}%
\bibitem [{\citenamefont {Batista}\ and\ \citenamefont
  {Aligia}(2004)}]{batista}%
  \BibitemOpen
  \bibfield  {author} {\bibinfo {author} {\bibfnamefont {C.~D.}\ \bibnamefont
  {Batista}}\ and\ \bibinfo {author} {\bibfnamefont {A.~A.}\ \bibnamefont
  {Aligia}},\ }\href {\doibase 10.1103/PhysRevLett.92.246405} {\bibfield
  {journal} {\bibinfo  {journal} {Phys. Rev. Lett.}\ }\textbf {\bibinfo
  {volume} {92}},\ \bibinfo {pages} {246405} (\bibinfo {year}
  {2004})}\BibitemShut {NoStop}%
\bibitem [{\citenamefont {Garg}\ \emph {et~al.}(2006)\citenamefont {Garg},
  \citenamefont {Krishnamurthy},\ and\ \citenamefont {Randeria}}]{randeria}%
  \BibitemOpen
  \bibfield  {author} {\bibinfo {author} {\bibfnamefont {A.}~\bibnamefont
  {Garg}}, \bibinfo {author} {\bibfnamefont {H.~R.}\ \bibnamefont
  {Krishnamurthy}}, \ and\ \bibinfo {author} {\bibfnamefont {M.}~\bibnamefont
  {Randeria}},\ }\href {\doibase 10.1103/PhysRevLett.97.046403} {\bibfield
  {journal} {\bibinfo  {journal} {Phys. Rev. Lett.}\ }\textbf {\bibinfo
  {volume} {97}},\ \bibinfo {pages} {046403} (\bibinfo {year}
  {2006})}\BibitemShut {NoStop}%
\bibitem [{\citenamefont {Paris}\ \emph {et~al.}(2007)\citenamefont {Paris},
  \citenamefont {Bouadim}, \citenamefont {Hebert}, \citenamefont {Batrouni},\
  and\ \citenamefont {Scalettar}}]{scalettar}%
  \BibitemOpen
  \bibfield  {author} {\bibinfo {author} {\bibfnamefont {N.}~\bibnamefont
  {Paris}}, \bibinfo {author} {\bibfnamefont {K.}~\bibnamefont {Bouadim}},
  \bibinfo {author} {\bibfnamefont {F.}~\bibnamefont {Hebert}}, \bibinfo
  {author} {\bibfnamefont {G.~G.}\ \bibnamefont {Batrouni}}, \ and\ \bibinfo
  {author} {\bibfnamefont {R.~T.}\ \bibnamefont {Scalettar}},\ }\href {\doibase
  10.1103/PhysRevLett.98.046403} {\bibfield  {journal} {\bibinfo  {journal}
  {Phys. Rev. Lett.}\ }\textbf {\bibinfo {volume} {98}},\ \bibinfo {pages}
  {046403} (\bibinfo {year} {2007})}\BibitemShut {NoStop}%
\bibitem [{\citenamefont {Moeller}\ \emph {et~al.}(1999)\citenamefont
  {Moeller}, \citenamefont {Dobrosavljevi\ifmmode~\acute{c}\else \'{c}\fi{}},\
  and\ \citenamefont {Ruckenstein}}]{ruckenstein}%
  \BibitemOpen
  \bibfield  {author} {\bibinfo {author} {\bibfnamefont {G.}~\bibnamefont
  {Moeller}}, \bibinfo {author} {\bibfnamefont {V.}~\bibnamefont
  {Dobrosavljevi\ifmmode~\acute{c}\else \'{c}\fi{}}}, \ and\ \bibinfo {author}
  {\bibfnamefont {A.~E.}\ \bibnamefont {Ruckenstein}},\ }\href {\doibase
  10.1103/PhysRevB.59.6846} {\bibfield  {journal} {\bibinfo  {journal} {Phys.
  Rev. B}\ }\textbf {\bibinfo {volume} {59}},\ \bibinfo {pages} {6846}
  (\bibinfo {year} {1999})}\BibitemShut {NoStop}%
\bibitem [{\citenamefont {Fuhrmann}\ \emph {et~al.}(2006)\citenamefont
  {Fuhrmann}, \citenamefont {Heilmann},\ and\ \citenamefont {Monien}}]{monien}%
  \BibitemOpen
  \bibfield  {author} {\bibinfo {author} {\bibfnamefont {A.}~\bibnamefont
  {Fuhrmann}}, \bibinfo {author} {\bibfnamefont {D.}~\bibnamefont {Heilmann}},
  \ and\ \bibinfo {author} {\bibfnamefont {H.}~\bibnamefont {Monien}},\ }\href
  {\doibase 10.1103/PhysRevB.73.245118} {\bibfield  {journal} {\bibinfo
  {journal} {Phys. Rev. B}\ }\textbf {\bibinfo {volume} {73}},\ \bibinfo
  {pages} {245118} (\bibinfo {year} {2006})}\BibitemShut {NoStop}%
\bibitem [{\citenamefont {Kancharla}\ and\ \citenamefont
  {Okamoto}(2007)}]{okamoto}%
  \BibitemOpen
  \bibfield  {author} {\bibinfo {author} {\bibfnamefont {S.~S.}\ \bibnamefont
  {Kancharla}}\ and\ \bibinfo {author} {\bibfnamefont {S.}~\bibnamefont
  {Okamoto}},\ }\href {\doibase 10.1103/PhysRevB.75.193103} {\bibfield
  {journal} {\bibinfo  {journal} {Phys. Rev. B}\ }\textbf {\bibinfo {volume}
  {75}},\ \bibinfo {pages} {193103} (\bibinfo {year} {2007})}\BibitemShut
  {NoStop}%
\bibitem [{\citenamefont {Sentef}\ \emph {et~al.}(2009)\citenamefont {Sentef},
  \citenamefont {Kune\ifmmode~\check{s}\else \v{s}\fi{}}, \citenamefont
  {Werner},\ and\ \citenamefont {Kampf}}]{kampf}%
  \BibitemOpen
  \bibfield  {author} {\bibinfo {author} {\bibfnamefont {M.}~\bibnamefont
  {Sentef}}, \bibinfo {author} {\bibfnamefont {J.}~\bibnamefont
  {Kune\ifmmode~\check{s}\else \v{s}\fi{}}}, \bibinfo {author} {\bibfnamefont
  {P.}~\bibnamefont {Werner}}, \ and\ \bibinfo {author} {\bibfnamefont {A.~P.}\
  \bibnamefont {Kampf}},\ }\href {\doibase 10.1103/PhysRevB.80.155116}
  {\bibfield  {journal} {\bibinfo  {journal} {Phys. Rev. B}\ }\textbf {\bibinfo
  {volume} {80}},\ \bibinfo {pages} {155116} (\bibinfo {year}
  {2009})}\BibitemShut {NoStop}%
\bibitem [{\citenamefont {Lee}\ \emph {et~al.}(2014)\citenamefont {Lee},
  \citenamefont {Zhang}, \citenamefont {Jeschke},\ and\ \citenamefont
  {Valent\'{\i}}}]{valenti}%
  \BibitemOpen
  \bibfield  {author} {\bibinfo {author} {\bibfnamefont {H.}~\bibnamefont
  {Lee}}, \bibinfo {author} {\bibfnamefont {Y.-Z.}\ \bibnamefont {Zhang}},
  \bibinfo {author} {\bibfnamefont {H.~O.}\ \bibnamefont {Jeschke}}, \ and\
  \bibinfo {author} {\bibfnamefont {R.}~\bibnamefont {Valent\'{\i}}},\ }\href
  {\doibase 10.1103/PhysRevB.89.035139} {\bibfield  {journal} {\bibinfo
  {journal} {Phys. Rev. B}\ }\textbf {\bibinfo {volume} {89}},\ \bibinfo
  {pages} {035139} (\bibinfo {year} {2014})}\BibitemShut {NoStop}%
\bibitem [{\citenamefont {Gall}\ \emph {et~al.}(2021)\citenamefont {Gall},
  \citenamefont {Wurz}, \citenamefont {Samland}, \citenamefont {Chan},\ and\
  \citenamefont {K{\"o}hl}}]{gall2021}%
  \BibitemOpen
  \bibfield  {author} {\bibinfo {author} {\bibfnamefont {M.}~\bibnamefont
  {Gall}}, \bibinfo {author} {\bibfnamefont {N.}~\bibnamefont {Wurz}}, \bibinfo
  {author} {\bibfnamefont {J.}~\bibnamefont {Samland}}, \bibinfo {author}
  {\bibfnamefont {C.~F.}\ \bibnamefont {Chan}}, \ and\ \bibinfo {author}
  {\bibfnamefont {M.}~\bibnamefont {K{\"o}hl}},\ }\href {\doibase
  10.1038/s41586-020-03058-x} {\bibfield  {journal} {\bibinfo  {journal}
  {Nature}\ }\textbf {\bibinfo {volume} {589}},\ \bibinfo {pages} {40}
  (\bibinfo {year} {2021})}\BibitemShut {NoStop}%
\bibitem [{\citenamefont {Gull}\ \emph {et~al.}(2011)\citenamefont {Gull},
  \citenamefont {Millis}, \citenamefont {Lichtenstein}, \citenamefont
  {Rubtsov}, \citenamefont {Troyer},\ and\ \citenamefont {Werner}}]{gull}%
  \BibitemOpen
  \bibfield  {author} {\bibinfo {author} {\bibfnamefont {E.}~\bibnamefont
  {Gull}}, \bibinfo {author} {\bibfnamefont {A.~J.}\ \bibnamefont {Millis}},
  \bibinfo {author} {\bibfnamefont {A.~I.}\ \bibnamefont {Lichtenstein}},
  \bibinfo {author} {\bibfnamefont {A.~N.}\ \bibnamefont {Rubtsov}}, \bibinfo
  {author} {\bibfnamefont {M.}~\bibnamefont {Troyer}}, \ and\ \bibinfo {author}
  {\bibfnamefont {P.}~\bibnamefont {Werner}},\ }\href {\doibase
  10.1103/RevModPhys.83.349} {\bibfield  {journal} {\bibinfo  {journal} {Rev.
  Mod. Phys.}\ }\textbf {\bibinfo {volume} {83}},\ \bibinfo {pages} {349}
  (\bibinfo {year} {2011})}\BibitemShut {NoStop}%
\bibitem [{\citenamefont {Huang}\ \emph {et~al.}(2015)\citenamefont {Huang},
  \citenamefont {Wang}, \citenamefont {Meng}, \citenamefont {Du}, \citenamefont
  {Werner},\ and\ \citenamefont {Dai}}]{iqist}%
  \BibitemOpen
  \bibfield  {author} {\bibinfo {author} {\bibfnamefont {L.}~\bibnamefont
  {Huang}}, \bibinfo {author} {\bibfnamefont {Y.}~\bibnamefont {Wang}},
  \bibinfo {author} {\bibfnamefont {Z.~Y.}\ \bibnamefont {Meng}}, \bibinfo
  {author} {\bibfnamefont {L.}~\bibnamefont {Du}}, \bibinfo {author}
  {\bibfnamefont {P.}~\bibnamefont {Werner}}, \ and\ \bibinfo {author}
  {\bibfnamefont {X.}~\bibnamefont {Dai}},\ }\href {\doibase
  https://doi.org/10.1016/j.cpc.2015.04.020} {\bibfield  {journal} {\bibinfo
  {journal} {Computer Physics Communications}\ }\textbf {\bibinfo {volume}
  {195}},\ \bibinfo {pages} {140} (\bibinfo {year} {2015})}\BibitemShut
  {NoStop}%
\end{thebibliography}%


\begin{thebibliography}{5}%
\expandafter\ifx\csname natexlab\endcsname\relax\def\natexlab#1{#1}\fi
\expandafter\ifx\csname bibnamefont\endcsname\relax
  \def\bibnamefont#1{#1}\fi
\expandafter\ifx\csname bibfnamefont\endcsname\relax
  \def\bibfnamefont#1{#1}\fi
\expandafter\ifx\csname citenamefont\endcsname\relax
  \def\citenamefont#1{#1}\fi
\expandafter\ifx\csname url\endcsname\relax
  \def\url#1{\texttt{#1}}\fi
\expandafter\ifx\csname urlprefix\endcsname\relax\def\urlprefix{URL }\fi
\providecommand{\bibinfo}[2]{#2}
\providecommand{\eprint}[2][]{\url{#2}}
\let\auto@bib@innerbib\@empty
\bibitem[{\citenamefont{Kresse and Furthm{\"u}ller}(1996)}]{kresse1996}
\bibinfo{author}{\bibfnamefont{G.}~\bibnamefont{Kresse}} \bibnamefont{and}
  \bibinfo{author}{\bibfnamefont{J.}~\bibnamefont{Furthm{\"u}ller}},
  \bibinfo{journal}{Comput. Mater. Sci.} \textbf{\bibinfo{volume}{6}},
  \bibinfo{pages}{15} (\bibinfo{year}{1996}).

\bibitem[{\citenamefont{Kresse and Joubert}(1999)}]{Joubert1999}
\bibinfo{author}{\bibfnamefont{G.}~\bibnamefont{Kresse}} \bibnamefont{and}
  \bibinfo{author}{\bibfnamefont{D.}~\bibnamefont{Joubert}},
  \bibinfo{journal}{Phys. Rev. B} \textbf{\bibinfo{volume}{59}},
  \bibinfo{pages}{1758} (\bibinfo{year}{1999}).

\bibitem[{\citenamefont{Perdew et~al.}(1996)\citenamefont{Perdew, Burke, and
  Ernzerhof}}]{perdew1996}
\bibinfo{author}{\bibfnamefont{J.~P.} \bibnamefont{Perdew}},
  \bibinfo{author}{\bibfnamefont{K.}~\bibnamefont{Burke}}, \bibnamefont{and}
  \bibinfo{author}{\bibfnamefont{M.}~\bibnamefont{Ernzerhof}},
  \bibinfo{journal}{Phys. Rev. Lett.} \textbf{\bibinfo{volume}{77}},
  \bibinfo{pages}{3865} (\bibinfo{year}{1996}).
  
\bibitem[{\citenamefont{{Gao} et~al.}(2022)\citenamefont{{Gao}, {Zhang},
  {Wang}, {Tao}, {Liu}, {Wang}, {Yuan}, {Qu}, {Pan}, {Peng}
  et~al.}}]{arpes_supp}
\bibinfo{author}{\bibfnamefont{S.}~\bibnamefont{{Gao}}},
  \bibinfo{author}{\bibfnamefont{S.}~\bibnamefont{{Zhang}}},
  \bibinfo{author}{\bibfnamefont{C.}~\bibnamefont{{Wang}}},
  \bibinfo{author}{\bibfnamefont{W.}~\bibnamefont{{Tao}}},
  \bibinfo{author}{\bibfnamefont{J.}~\bibnamefont{{Liu}}},
  \bibinfo{author}{\bibfnamefont{T.}~\bibnamefont{{Wang}}},
  \bibinfo{author}{\bibfnamefont{S.}~\bibnamefont{{Yuan}}},
  \bibinfo{author}{\bibfnamefont{G.}~\bibnamefont{{Qu}}},
  \bibinfo{author}{\bibfnamefont{M.}~\bibnamefont{{Pan}}},
  \bibinfo{author}{\bibfnamefont{S.}~\bibnamefont{{Peng}}},
  \bibnamefont{et~al.}, \bibinfo{journal}{arXiv e-prints}
  \bibinfo{eid}{arXiv:2205.11462} (\bibinfo{year}{2022}), \eprint{2205.11462}.

\bibitem[{\citenamefont{Zhou and Wang}(2008)}]{zhouwang07_supp}
\bibinfo{author}{\bibfnamefont{S.}~\bibnamefont{Zhou}} \bibnamefont{and}
  \bibinfo{author}{\bibfnamefont{Z.}~\bibnamefont{Wang}},
  \bibinfo{journal}{Phys. Rev. Lett.} \textbf{\bibinfo{volume}{100}},
  \bibinfo{pages}{217002} (\bibinfo{year}{2008}).

\end{thebibliography}

\clearpage
\onecolumngrid
\begin{center}
\textbf{\large Supplemental Material: Mottness in two-dimensional van der Waals Nb$_3$X$_8$ monolayers (X=Cl, Br, and I)}
\end{center}

\setcounter{equation}{0}
\setcounter{figure}{0}
\setcounter{table}{0}
\setcounter{page}{1}
\makeatletter
\renewcommand{\theequation}{S\arabic{equation}}
\renewcommand{\thefigure}{S\arabic{figure}}
\renewcommand{\bibnumfmt}[1]{[S#1]}
\renewcommand{\citenumfont}[1]{S#1}

\onecolumngrid

\section{Computational details of the first-principle calculation}
\begin{figure}[h]
	\begin{center}
		\fig{3.4in}{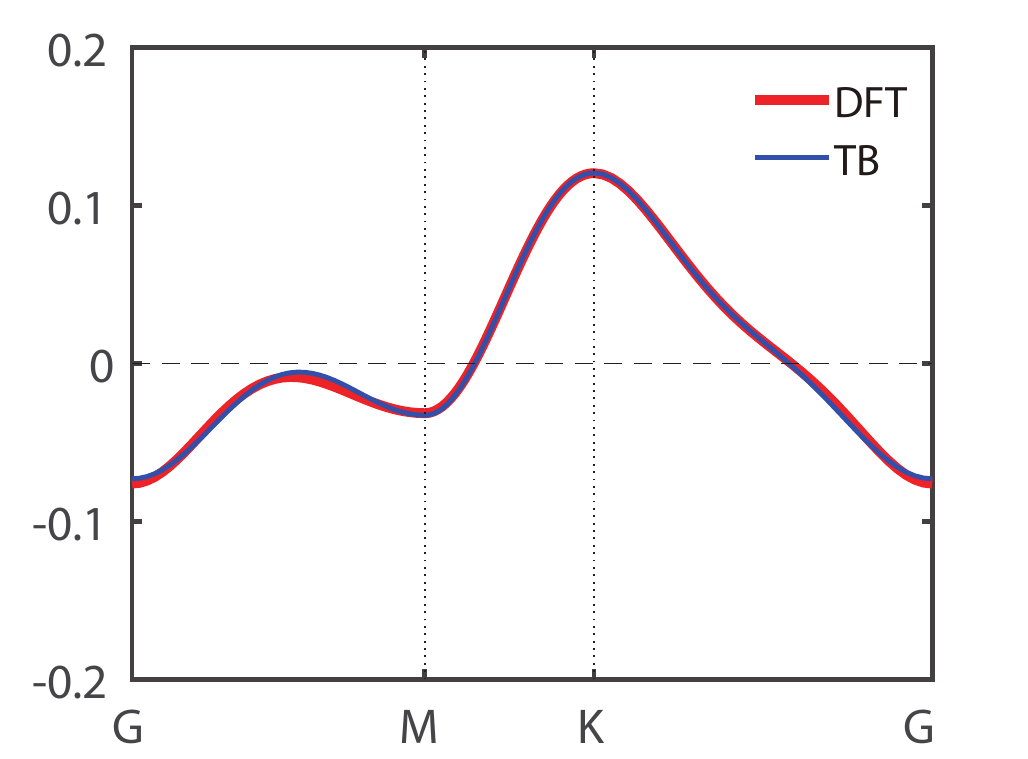}\caption{The monolayer band structure of Nb$_3$I$_8$ with DFT calculation (red line) and Wannierization fitting TB (blue line) with the bandwidth around 0.196 eV.
			\label{fig:NbI}}
	\end{center}
	\vskip-0.5cm
\end{figure}

Our density functional theory (DFT) calculation is performed by Vienna ab initio simulation package (VASP) code~\cite{kresse1996} with the projector augmented wave (PAW) method~\cite{Joubert1999}. The Perdew-Burke-Ernzerhof~(PBE)~\cite{perdew1996} exchange-correlation functional is used in our calculation. The kinetic energy cutoff is set to be 600 eV for the expanding the wave functions into a plane-wave basis. The energy convergence criterion is $10^{-7}$ eV and the $\Gamma$-centered~\textbf{k}-mesh is $12\times12\times2$. The monolayer Nb$_3$X$_8$ (X=Cl,Br,I) is fully relaxed with built-in 40 \AA$ $ thick vacuum layer while forces are minimized to less than 0.001 eV/{\AA}. The hopping parameters of tight-binding models are fitted from DFT-calculated band structures of monolayer Nb$_3$X$_8$. The band structure of Nb$_3$Cl$_8$ and Nb$_3$Br$_8$ are already shown in Fig.1 of the main text and the band structure of Nb$_3$I$_8$ is shown in Fig.~\ref{fig:NbI}.

\section{Local interaction problem for the bilayer Hubbard model}
The crossover between band insulator and Mott insulator can be understood from the local site problem of two electrons.
The local Hamiltonian for the single site model can be written as 
\begin{equation}
	\hat{H}_{loc}=\left(\begin{array}{cccccc}
		0\\
		& U_{v} & -J\\
		& -J & U_{v}\\
		&  &  & 0\\
		&  &  &  & -2t_{\perp}+U_{0} & J\\
		&  &  &  & J & 2t_{\perp}+U_{0}
	\end{array}\right)\begin{array}{c}
		\left|\uparrow,\uparrow\right\rangle _{a}\\
		\left|\uparrow,\downarrow\right\rangle _{a}\\
		\left|\downarrow,\uparrow\right\rangle _{a}\\
		\left|\downarrow,\downarrow\right\rangle _{a}\\
		\left|\uparrow\downarrow,0\right\rangle _{a}\\
		\left|0,\uparrow\downarrow\right\rangle _{a}
	\end{array}
\end{equation}
in the 2-electron band basis $|\uparrow, \uparrow\rangle_a$, $|\uparrow, \downarrow\rangle_a$, $|\downarrow, \uparrow\rangle_a$, $|\downarrow, \downarrow\rangle_a$, $|\uparrow\downarrow, 0\rangle_a$, and $|0,\uparrow\downarrow\rangle_a$,
which is defined as
\begin{equation}
	\left\{ \begin{array}{c}
		\\
		\\
		\\
	\end{array}\right.\begin{array}{c}
		\left|\sigma,\sigma^{\prime}\right\rangle _{a}=a_{+\sigma^{\prime}}^{\dagger}a_{-\sigma}^{\dagger}\left|0\right>\\
		\left|\uparrow\downarrow,0\right\rangle _{a}=a_{-\downarrow}^{\dagger}a_{-\uparrow}^{\dagger}\left|0\right>\\
		\left|0,\uparrow\downarrow\right\rangle _{a}=a_{+\downarrow}^{\dagger}a_{+\uparrow}^{\dagger}\left|0\right>
	\end{array}
\end{equation}
Here, $U_v=J=U_0=\frac{U}{2}$.
The local interaction model in the original basis is 
\begin{eqnarray}
	\hat{H}_{loc}=\left(\begin{array}{cccccc} 
		0 &  &  & & &\\
		& 0 &  & & t_{\perp} & t_{\perp}\\
		&  &  0 & & -t_{\perp} & -t_{\perp} \\
		& &  & 0& &\\
		& t_{\perp} & -t_{\perp}  & & U & 0\\
		& t_{\perp} & -t_{\perp} & & 0 & U
	\end{array}\right)\begin{array}{c}
		\left|\uparrow,\uparrow\right\rangle _{c}\\
		\left|\uparrow,\downarrow\right\rangle _{c}\\
		\left|\downarrow,\uparrow\right\rangle _{c}\\
		\left|\downarrow,\downarrow\right\rangle _{c}\\
		\left|\uparrow\downarrow,0\right\rangle _{c}\\
		\left|0,\uparrow\downarrow\right\rangle _{c}
	\end{array}
\end{eqnarray}
in the 2-electron orbital basis $|\uparrow, \uparrow\rangle_c$, $|\uparrow, \downarrow\rangle_c$, $|\downarrow, \uparrow\rangle_c$, $|\downarrow, \downarrow\rangle_c$, $|\uparrow\downarrow, 0\rangle_c$, and $|0,\uparrow\downarrow\rangle_c$ defined as
\begin{equation}
	\left\{ \begin{array}{c}
		\\
		\\
		\\
	\end{array}\right.\begin{array}{c}
		\left|\sigma,\sigma^{\prime}\right\rangle _{c}=c_{2\sigma^{\prime}}^{\dagger}c_{1\sigma}^{\dagger}\left|0\right>\\
		\left|\uparrow\downarrow,0\right\rangle _{c}=c_{1\downarrow}^{\dagger}c_{1\uparrow}^{\dagger}\left|0\right>\\
		\left|0,\uparrow\downarrow\right\rangle _{c}=c_{2\downarrow}^{\dagger}c_{2\uparrow}^{\dagger}\left|0\right>
	\end{array}
\end{equation}
Diagonalizing $\hat{H}_{loc}$ leads to the eigenvalues and eigenstates in TABLE I.

To see the difference between band and Mott insulator, we study the properties of the retarded Green's function in the two limiting cases.
Deep in the Mott insulator case, we consider the atomic limit. In the zero temperature limit, the Lehmann representation of the Green's function leads to the local Green's function as
\begin{equation}
	G^R_{\alpha\sigma,\alpha^{\prime}\sigma^{\prime}}(\omega)=\sum_n \left( \frac{\left\langle 0|a_{\alpha\sigma}|n \right\rangle \left\langle n|a^{\dagger}_{\alpha^{\prime}\sigma^{\prime}}|0 \right\rangle}{\omega-\omega_{n0}+i\eta} + \frac{\left\langle 0|a^{\dagger}_{\alpha^{\prime}\sigma^{\prime}}|n \right\rangle \left\langle n|a_{\alpha\sigma}|0 \right\rangle}{\omega+\omega_{n0}+i\eta} \right)
\end{equation}
where, $|n\rangle$ runs over all the eigenstates of the local Hamiltonian $\hat H_{loc}-\mu \hat N$, $|0\rangle$ is the ground state  corresponding to $|\Gamma_1\rangle$ listed in TABLE I of the main text for the case with half filling and $\omega_{n0}$ corresponds to the energy difference between states $|n\rangle$ and $|0\rangle$ and $\eta$ is an infinitesimal positive number.
Here since $|0\rangle$ contains two electrons, $|n\rangle$ has to be the states with either one or three electrons. The local Hamiltonian in the 1-electron and 3-electron sectors are 
\begin{equation}
	\hat{H}_{loc}=\left(\begin{array}{cccc}
		-t_{\perp}\\
		& -t_{\perp} \\
		&  & t_{\perp}\\
		&  &  & t_{\perp}
	\end{array}\right)\begin{array}{c}
		\left|\uparrow,0\right\rangle _{a}\\
		\left|\downarrow,0\right\rangle _{a}\\
		\left|0,\uparrow\right\rangle _{a}\\
		\left|0,\downarrow\right\rangle _{a}
	\end{array}
\end{equation}
and 
\begin{equation}
	\hat{H}_{loc}=\left(\begin{array}{cccc}
		U_0+U_v-t_{\perp}\\
		& U_0+U_v-t_{\perp} \\
		&  & U_0+U_v+t_{\perp}\\
		&  &  & U_0+U_v+t_{\perp}
	\end{array}\right)\begin{array}{c}
		\left|\uparrow\downarrow,\uparrow\right\rangle _{a}\\
		\left|\uparrow\downarrow,\downarrow\right\rangle _{a}\\
		\left|\uparrow,\uparrow\downarrow\right\rangle _{a}\\
		\left|\downarrow,\uparrow\downarrow\right\rangle _{a}
	\end{array}\ .
\end{equation}
The calculated Green's function is diagonal in both spin and band basis and can be written  as
\begin{equation}
	G^{R}_{\alpha\sigma,\alpha\sigma}(\omega)=\frac{1}{1+\zeta^2}\left[ \frac{1}{\omega-(\bar\alpha t_{\perp}+E_{\alpha}-\mu)+i\eta}+\frac{\zeta^2}{\omega-(\bar\alpha t_{\perp}+E_{\bar\alpha}-\mu)+i\eta} \right]
\end{equation}
where, $\zeta=\frac{U}{\sqrt{16t_{\perp}^2+U^2}+4t_{\perp}}$ and $E_{\pm}=\frac{1}{2}(U\pm\sqrt{16t_{\perp}^2+U^2})$.
Clearly, the Green's functions have four poles located at $\pm t_{\perp}+E_{\pm}$ for energy measured from the chemical potential $\mu$, so that sufficient larger interlayer hopping $t_{\perp}$ can split the two Hubbard bands into four which is confirmed by the ARPES experiment~\cite{arpes_supp}.
On the other hand, in the band insulator limit with vanishing U, the Green's function is easy to get which can be written as
\begin{equation}
 	G^{R}_{\alpha\sigma,\alpha\sigma}(\omega)=\frac{1}{\omega-(\varepsilon_{k}+\alpha t_{\perp}-\mu)+i\eta}
\end{equation}
which only have two poles located at $\varepsilon_k\pm\alpha t_{\perp}$ for energy measured from $\mu$.
The difference in the number of poles of the Green's function can serve as an experimental feature to differentiate the two insulating states.

\section{Slave boson approach}
In the Kotliar and Ruckenstein slave boson method, the local Hilbert space is represented by a spin-1/2 fermion $f_{\sigma}$ and four slave bosons: $e$ (holon), $d$ (doublon), and $p_{\sigma}$, so that the empty state $\left|0\right>=e^{\dagger}\left|vac\right>$, the singly occupied state $\left|\sigma\right>=p_{\sigma}^{\dagger}f_{\sigma}^{\dagger}\left|vac\right>$ and the doubly occupied state $\left|\uparrow\downarrow\right>=d^{\dagger}f_{\downarrow}^{\dagger}f_{\uparrow}^{\dagger}\left|vac\right>$.
The completeness of the local Hilbert space requires that the slave bosons satisfy the local constraints as,
\begin{equation}
	Q_{i}=e_{i}^{\dagger}e_{i}+\sum_{\sigma}p_{i,\sigma}^{\dagger} p_{i,\sigma}+d_{i,\sigma}^{\dagger}d_{i,\sigma}-1=0
	\label{eq:cons1}
\end{equation}
and the equivalence between the fermion and boson representation of the particle density requires the further constraint as,
\begin{equation}
	Q_{i,\sigma}=p_{i,\sigma}^{\dagger} p_{i,\sigma}+d_{i}^{\dagger}d_{i}-f_{i,\sigma}^{\dagger} f_{i,\sigma}=0 \ ,
	\label{eq:cons2}
\end{equation} 
Then the Hubbard model can be faithfully written as

\begin{equation}
		H= \sum_{ij\sigma} t_{ij} g_{i,\sigma}^{\dagger} g_{j,\sigma} f_{i,\sigma}^{\dagger} f_{j,\sigma}  +h.c. 
		+ U\sum_{i} d_{i}d_{i} - \mu\sum_{i}f_{i,\sigma}^{\dagger} f_{i,\sigma}-\sum_{i}\alpha_{i}Q_{i}-\sum_{i}\lambda_{i,\sigma}Q_{i,\sigma}
	\label{eq:sb}
\end{equation}
where the $\alpha_{i}$ and $\lambda_{i,\sigma}$ are Lagrange multipliers to enforce the constraints and $g_{i,\sigma}$ are the renormalization factors introduced for the hopping terms as
\begin{equation}
	g_{i,\sigma}= L_{i,\sigma}^{-\frac{1}{2}} \left( e_{i}^{\dagger}p_{i,\sigma}+p_{i,\bar\sigma}^{\dagger}d_{i} \right) R_{i,\bar\sigma}^{-\frac{1}{2}}
\end{equation}  
with $L_{i,\sigma}= 1-d_{i}^{\dagger}d_{i}-p_{i,\sigma}^{\dagger} p_{i,\sigma}$ and
$R_{i,\sigma}= 1-e_{i}^{\dagger}e_{i}-p_{i,\sigma}^{\dagger} p_{i,\sigma}$.
The mean-field solution corresponds to condensing all the boson fields uniformly which are determined self consistently by minimizing the ground state energy $<H>$ with respect to the condensed boson fields as well as the Lagrange multipliers ($e_{i}$, $d_{i}$, $p_{i,\sigma}$, $\alpha_{i}$, $\lambda_{i,\sigma}$)=($e$, $d$, $p_{\sigma}$, $\alpha$, $\lambda_{\sigma}$). For the uniformly condensed boson fields, the Hamiltonian Eq.~\ref{eq:sb} can be written in momentum space as
\begin{equation}
		H= \sum_{\bk\sigma} \varepsilon_{\bk} g_{\sigma}^2 f_{\bk,\sigma}^{\dagger} f_{\bk,\sigma}  +h.c.
	- (\mu+\lambda_{\sigma})\sum_{\bk}f_{\bk,\sigma}^{\dagger} f_{\bk,\sigma} + N_sU d^2
		-N_s\lambda_{\sigma}(p_{\sigma}^{2}+d^{2}) -N_s\alpha(e^{2}+\sum_{\sigma}p_{\sigma}^{2}+d_{\sigma}^{2}-1)
\end{equation}
where
\begin{equation}
	\begin{split}
		\varepsilon_{\bk}&=2t_{1}\left(\cos(k_1)+\cos(k_2)+\cos(k_3)\right)
		+2t_{2}\left(\cos(k_1+k_2)+\cos(k_2+k_3)+\cos(k_3-k_1)\right)
		\\
		&+2t_{3}\left(\cos(2k_1)+\cos(2k_2)+\cos(2k_3)\right) \ ,
	\end{split}
\end{equation}
with $k_1=k_x$, $k_2=\frac{1}{2}k_x+\frac{\sqrt{3}}{2}k_y$, $k_3=-\frac{1}{2}k_x+\frac{\sqrt{3}}{2}k_y$ 
and
\begin{equation}
	g_{\sigma}=\left(1-d^{2}-p_{\sigma}^{2}\right)^{-\frac{1}{2}} \left( e p_{\sigma}+p_{\bar\sigma}d \right) \left(1-e^{2}-p_{\bar\sigma}^{2}\right)^{-\frac{1}{2}} \ .
\end{equation}
We solve for the boson field numerically in a mesh of 600$\times$600 k points via minimizing the ground state energy $\left<H\right>$ with respect to ($e$, $d$, $p_{\sigma}$, $\alpha$, $\lambda_{\sigma}$)  together
with the chemical potential $\mu$ determined from the average particle density $n=(1/N_s)\sum_{i,\sigma}\left<f_{i,\sigma}^{\dagger}f_{i,\sigma}\right>=1-x$. The determined phase diagrams for nonmagnetic and magnetic solutions are shown in Fig.2\textbf{a,b} of the main text.

\section{Topological property of the  \lowercase{d+id} superconductor}
As mentioned in the main text that the $d_{x^2-y^2}+id_{xy}$ pair breaks the time-reversal symmetry which can lead to possible non-trivial topological property. Here we take the $d_{x^2-y^2}+id_{xy}$ pairing symmetry with the next nearest neighbor pairing for the monolayer system as an example to show the non-trivial topological property. 
The total Hamiltonian of the superconductor can be written as
\begin{equation}
	H=\sum_{k\sigma}(\varepsilon_k-\mu)c_{k\sigma}^{\dagger}c_{k\sigma}+\sum_k (\Delta_k c_{k\uparrow}^{\dagger}c_{-k\downarrow}^{\dagger}+h.c.)=\sum_k h_k
\end{equation} 
For the next nearest neighbor $d_{x^2-y^2}+id_{xy}$ pairing, 
the pairing potential has the form $\Delta_k=\Delta_k^{\pm}=\Delta_0\left[\gamma_1(k)\pm i\gamma_2(k)\right]$
with
\begin{equation}
  \gamma_1(k)=2\cos(k_2+k_3)-\cos(k_1+k_2)-\cos(k_1-k_3)
\end{equation}
\begin{equation}
	\gamma_2(k)=-\sqrt{3}\left[\cos(k_1-k_3)-\cos(k_2+k_3)\right]
\end{equation}
with $k_1=k_x$, $k_2=\frac{1}{2}k_x+\frac{\sqrt{3}}{2}k_y$, $k_3=-\frac{1}{2}k_x+\frac{\sqrt{3}}{2}k_y$, 
which satisfies the relation $\Delta_{C_3(k)}^{\pm}=e^{\pm i\frac{2\pi}{3}}\Delta_k^{\pm}$ under the $C_3$ rotation. 
To determine the topological property of the superconductor, we calculate the Berry curvature of the system as
\begin{equation}
	\label{eq:berry}
	{\bf\Omega}_n(k)=i\left<\nabla_k u_n(k)\right| \times \left| \nabla_k u_n(k) \right>
\end{equation}
where, $u_n(k)$ is the nth eigenstate of $h_k$, so that the Chern number can de determined as 
\begin{equation}
	C_n=\int \frac{dk}{2\pi} {\bf\Omega}_n(k)
\end{equation}

\begin{figure}[h]
	\begin{center}
		\fig{3.4in}{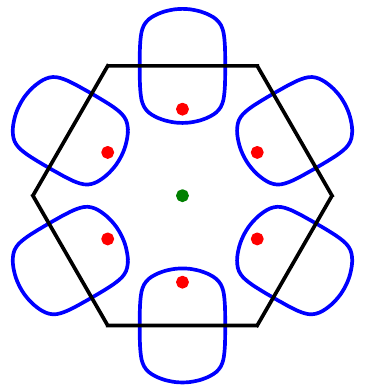}\caption{Location of the poles (red and green dots) of the pairing potential $\Delta_k^{\pm}$ together with the Fermi surface (blue lines) of the mono layer system with hole doping x=0.1. Here, the black hexagon corresponds to the Brillouin zone.
			\label{fig:pole}}
	\end{center}
	\vskip-0.5cm
\end{figure}
For the monolayer system with hole doping x=0.1, we find the Chern number $C=\pm6$ for $\Delta_k^{\pm}$ pairing.
This can be understood from the poles of the pairing potential $\Delta_k^{\pm}$ that are enclosed by the Fermi surface of the normal state system~\cite{zhouwang07_supp}. Here for the case we consider, the Fermi surface encloses six poles as shown in Fig~\ref{fig:pole}, each of which contributes $\pm2\pi$ of the Berry curvature, so that the total Chern number $C=\pm6$ for $\Delta_k^{\pm}$ pairing superconductor.

%


\end{document}